\documentclass[journal,draftcls,onecolumn,letterpaper,twoside,11pt]{IEEEtran}

\usepackage[cmex10]{amsmath}
\usepackage{latexsym,amssymb,cite,bbm,epsf,amsthm,subfigure,algorithm,algorithmic}
\usepackage{graphicx}
\usepackage{enumerate}
\usepackage{mathrsfs}
\usepackage{epstopdf}
\usepackage{bbm}
\usepackage{color}

\newcommand{\bc}{\begin{center}}
\newcommand{\ec}{\end{center}}
\newcommand{\be}{\begin{equation}}
\newcommand{\ee}{\end{equation}}
\newcommand{\ba}{\begin{array}}
\newcommand{\ea}{\end{array}}
\newcommand{\bea}{\begin{eqnarray}}
\newcommand{\eea}{\end{eqnarray}}
\newcommand{\bal}{\begin{align}}
\newcommand{\eal}{\end{align}}
\newcommand{\ei}{\end{itemize}}
\newcommand{\bi}{\begin{itemize}}
\newcommand{\bfi}{\begin{figure}}
\newcommand{\efi}{\end{figure}}
\newcommand{\MB}{\left[\begin{array}}
\newcommand{\ME}{\end{array}\right]}
\newcommand{\nn}{\nonumber}

\newtheorem{thm}{Theorem}

\newtheorem{lem}{Lemma}
\newtheorem{pro}{Proposition}

\renewcommand{\vec}[1]{\mbox{\boldmath${#1}$}}
\newcommand{\mH}{\vec{H}}
\newcommand{\mI}{\vec{I}}
\newcommand{\mP}{\vec{P}}
\newcommand{\mU}{\vec{U}}
\newcommand{\mD}{\vec{D}}

\newcommand{\mR}{\vec{R}}

\newcommand{\Exp}{\mathsf{E}}
\newcommand{\Var}{\mathsf{Var}}
\newcommand{\Cov}{\mathsf{Cov}}

\newcommand{\Pro}{\mathsf{P}}

\newcommand{\Tr}{\mathsf{Tr}}
\newcommand{\x}{\mathsf{x}}
\newcommand{\X}{\mathsf{X}}

\newcommand{\cS}{\mathcal{S}}
\newcommand{\cT}{\mathcal{T}}
\newcommand{\cH}{\mathcal{H}}

\newcommand{\cN}{\mathcal{N}}

\newcommand{\cF}{\mathcal{F}}
\newcommand{\cV}{\mathcal{V}}
\newcommand{\cC}{\mathscr{C}}

\newcommand{\bN}{\mathbb{N}}
\newcommand{\bR}{\mathbb{R}}

\newcommand{\ind}[1]{\mathbbm{1}_{\{#1\}}}   

\newcommand{\ignore}[1]{{}}

\ignore{

\addtolength{\textfloatsep}{-.8\textfloatsep}
\addtolength{\floatsep}{-.8\floatsep}

\abovedisplayskip=0.2cm
\belowdisplayskip=0.2cm
\raggedbottom

\usepackage{tikz}

}

\begin{document}

\title{Sequential and Decentralized Estimation of Linear Regression Parameters in Wireless Sensor Networks}

\author{Yasin~Y{\i}lmaz\IEEEauthorrefmark{2}\footnote{\IEEEauthorrefmark{2}Electrical Engineering Department, Columbia University, New York, NY 10027.},\;\;
        George V. Moustakides\IEEEauthorrefmark{3}\footnote{\IEEEauthorrefmark{3}Dept. of Electrical \& Computer Engineering, University of Patras, 26500 Rio, Greece.},\;\;
        and \, Xiaodong Wang\IEEEauthorrefmark{2}}

\maketitle

\begin{abstract}
Sequential estimation of a vector of linear regression coefficients is considered under both centralized and decentralized setups. In sequential estimation, the number of observations used for estimation is determined by the observed samples, hence is random, as opposed to fixed-sample-size estimation. Specifically, after receiving a new sample, if a target accuracy level is reached, we stop and estimate using the samples collected so far; otherwise we continue to receive another sample. It is known that finding an optimum sequential estimator, which minimizes the average sample number for a given target accuracy level, is an intractable problem with a general stopping rule that depends on the complete observation history. By properly restricting the search space to stopping rules that depend on a specific subset of the complete observation history, we derive the optimum sequential estimator in the centralized case via optimal stopping theory. However, finding the optimum stopping rule in this case requires numerical computations that {\em quadratically} scales with the number of parameters to be estimated. For the decentralized setup with stringent energy constraints, under an alternative problem formulation that is conditional on the observed regressors, we first derive a simple optimum scheme whose computational complexity is {\em constant} with respect to the number of parameters. Then, following this simple optimum scheme we propose a decentralized sequential estimator whose computational complexity and energy consumption scales {\em linearly} with the number of parameters. 
Specifically, in the proposed decentralized scheme a close-to-optimum average stopping time performance is achieved by infrequently transmitting a single pulse with very short duration.
\end{abstract}


\section{Introduction}

In this paper, we are interested in sequentially estimating a vector of parameters (i.e., regression coefficients) $X\in\bR^n$ at a random stopping time $\cS$ in the following linear (regression) model,
\be
\label{eq:sig_mod}
    y_t = H_t^T X + w_t, ~t\in\bN,
\ee
where $y_t\in\bR$ is the observed sample, $H_t\in\bR^n$ is the vector of regressors and $w_t\in\bR$ is the additive noise. We consider the general case in which $H_t$ is random and observed at time $t$, which covers the deterministic $H_t$ case as a special case. This linear model is commonly used in many applications. For example, in system identification, $X$ is the unknown system coefficients, $H_t$ is the (random) input applied to the system, and $y_t$ is the output at time $t$. Another example is the estimation of wireless (multiple-access) channel coefficients, in which $X$ is the unknown channel coefficients, $H_t$ is the transmitted (random) pilot signal, $y_t$ is the received signal, and $w_t$ is the additive channel noise. 

Energy constraints are inherent to wireless sensor networks \cite{Akyildiz02}. Since data transmission is the primary source of energy consumption, it is essential to keep transmission rates low in wireless sensor networks, resulting in a {\em decentralized} setup. Decentralized parameter estimation is a fundamental task performed in wireless sensor networks \cite{Das09,Fang10,Ribeiro06,Msechu12,Xiao06,Xiao08,Luo05,Schizas07,Schizas08,Stankovic11,Borkar82,Zhao07}. 
In {\em sequential} estimation, the objective is to minimize the (average) number of observations for a given target accuracy level \cite{Ghosh97}. To that end, a sequential estimator $(\cS,\hat{X}_{\cS})$, as opposed to a traditional fixed-sample-size estimator, is equipped with a stopping rule which determines an appropriate time $\cS$ to stop taking new observations based on the observation history. Hence, the stopping time $\cS$ (i.e., the number of observations used in estimation) is a random variable. Endowed with a stopping mechanism, a sequential estimator saves not only time but also energy, both of which are critical resources. In particular, it avoids unnecessary data processing and transmission. 

Decentralized parameter estimation has been mainly studied under two different network topologies. In the first one, sensors communicate to a fusion center (FC) that performs estimation based on the received information, e.g., \cite{Fang10,Ribeiro06,Msechu12,Xiao08,Luo05,Schizas07}. The other commonly studied topology is called ad hoc network, in which there is no designated FC, but sensors compute their local estimators and communicate them through the network, e.g.,
\cite{Das09,Schizas08,Stankovic11,Borkar82,Zhao07}. Decentralized estimation under both network topologies is reviewed in \cite{Xiao06}. Many existing works consider parameter estimation in linear models, e.g., \cite{Das09,Fang10,Msechu12,Xiao08,Luo05,Stankovic11}. Whereas in \cite{Ribeiro06,Xiao06,Schizas07,Schizas08,Borkar82,Zhao07} a general nonlinear signal model is assumed. 
The majority of existing works on decentralized estimation, e.g., \cite{Das09,Fang10,Ribeiro06,Msechu12,Xiao06,Xiao08,Luo05,Schizas07,Schizas08,Stankovic11}, studies fixed-sample-size estimation. There are a few works, such as \cite{Borkar82,Fellouris12}, that consider sequential decentralized parameter estimation. Nevertheless, \cite{Borkar82} assumes that sensors transmit real numbers, and \cite{Fellouris12} focuses on continuous-time observations, which can be seen as practical limitations. 

In decentralized detection \cite{Fellouris11,Yilmaz12} and estimation \cite{Yilmaz14}, {\em level-triggered} sampling, an adaptive sampling technique which infrequently transmits a few bits, e.g., one bit, from sensors to the FC, has been used to achieve low-rate transmission. It has been also shown that the decentralized schemes based on level-triggered sampling significantly outperform their counterparts based on conventional uniform sampling in terms of average stopping time. 
We here propose a novel form of level-triggered sampling that infrequently transmits a single pulse from sensors to the FC, and at the same time achieves a close-to-optimum average stopping time performance. 

The stopping capability of sequential estimators comes with the cost of sophisticated analysis.
In most cases, it is not possible \textcolor{blue}{with discrete-time observations} to find an optimum sequential estimator that attains the sequential Cram\'er-Rao lower bound (CRLB) if the stopping time $\cS$ is adapted to the complete observation history \cite{Ghosh87}. Alternatively, in \cite{Grambsch83} and more recently in \cite{Fellouris12,Yilmaz14}, it was proposed to restrict $\cS$ to stopping times that are adapted to a specific subset of the complete observation history, which leads to simple optimum solutions. This idea of using a {\em restricted} stopping time first appeared in \cite{Grambsch83} with no optimality result. In \cite{Fellouris12}, with continuous-time observations, a sequential estimator with a restricted stopping time was shown to achieve the sequential version of the CRLB for scalar parameter estimation. In \cite{Yilmaz14}, \textcolor{blue}{for scalar parameter estimation} with discrete-time observations, a similar sequential estimator was shown to achieve the {\em conditional} sequential CRLB \textcolor{blue}{for the same restricted class of stopping times}. 

In this paper, for {\em vector} parameter estimation with discrete-time observations, we find the optimum sequential estimators that achieve the unconditional and conditional sequential CRLB for a certain class of stopping times. Moreover, we develop a computation- and energy-efficient decentralized scheme based on level-triggered sampling for sequential estimation of vector parameters. We should note here that the proposed vector parameter estimator is by no means a straightforward extension of the scalar parameter estimators in \cite{Grambsch83,Fellouris12,Yilmaz14}. Firstly, straightforward application of level-triggered sampling to the vector case yields a computational complexity and energy consumption that scale quadratically with the number of unknown parameters. We propose a linearly scaling method, which is analytically justified and numerically shown to perform close to the optimum average stopping time performance. Secondly, data transmission and thus energy consumption increase with the number of parameters, which may easily become prohibitive for a sensor with limited battery. We address this energy efficiency issue by infrequently transmitting a single pulse with very short duration, which encodes, in time, the overshoot in level-triggered sampling, achieving hence a close-to-optimum performance. 

The remainder of the paper is organized as follows. In Section \ref{sec:back}, we provide background information on linear parameter estimation. Then, in Section \ref{sec:opt}, for a restricted class of stopping rules that solely depend on the regressors $\{H_t\}$ in \eqref{eq:sig_mod}, we derive the optimum sequential estimators, that minimize the average sample number for a given target accuracy level, under two different formulations of the problem. Following the common practice in sequential analysis we first minimize the average stopping time subject to a constraint on the estimation accuracy which is a function of the estimator covariance. The optimum solution to this classical problem proves to be intractable for even moderate number of unknown parameters. Hence, it is not a convenient model for decentralized estimation. Therefore, we next follow an alternative approach and formulate the problem conditioned on the observed $\{H_t\}$ values, which yields a tractable optimum solution for any number of parameters. In Section \ref{sec:decent}, using the tractable solution of the conditional formulation as a model, we propose a computation- and energy-efficient decentralized sequential estimator based on level-triggered sampling. Finally, the paper is concluded in Section \ref{sec:conc}. 
We represent scalars with lower-case letters, vectors with upper-case letters and matrices with upper-case bold letters.

\section{Background}
\label{sec:back}

In \eqref{eq:sig_mod}, at each time $t$, we observe the sample $y_t$ and the vector $H_t$, hence $\{(y_{p},H_{p})\}_{p=1}^t$ are available. We assume $\{w_t\}$ are i.i.d. with $\Exp[w_t]=0$ and $\Var(w_t)=\sigma^2$. 
The least squares (LS) estimator minimizes the sum of squared errors, i.e.,
\be
\label{eq:LS}
    \hat{\X}_t = \arg \min_{X} \sum_{p=1}^t (y_{p}-H_{p}^T X)^2,
\ee
and is given by
\be
\label{eq:LS1}
    \hat{\X}_t = \left(\sum_{p=1}^t H_{p}H_{p}^T\right)^{-1} \sum_{p=1}^t H_{p}y_{p} = (\mH_t^T\mH_t)^{-1} \mH_t^T Y_t,
\ee
where $\mH_t=[H_1,\ldots,H_t]^T$ and $Y_t=[y_1,\ldots,y_t]^T$. Note that spatial diversity (i.e., a vector of observations and a regressor matrix at time $t$) can be easily incorporated in \eqref{eq:sig_mod} in the same way we deal with temporal diversity. Specifically, in \eqref{eq:LS} and \eqref{eq:LS1} we would also sum over the spatial dimensions.

Under the Gaussian noise, $w_t\sim\cN(0,\sigma^2)$, the LS estimator coincides with the minimum variance unbiased estimator (MVUE), and achieves the CRLB, i.e., $\Cov(\hat{X}_t|\mH_t)=\text{CRLB}_t$. To compute the CRLB we first write, given $X$ and $\mH_t$, the log-likelihood of the vector $Y_t$ as
\be
\label{eq:log-like}
    L_t = \log f(Y_t|X,\mH_t) = -\sum_{p=1}^t \frac{(y_{p}-H_{p}^T X)^2}{2\sigma^2} - \frac{t}{2}\log(2\pi\sigma^2).
\ee
Then,  we have
\be
\label{eq:CRLB}
	\text{CRLB}_t=\left(\Exp\left[ -\frac{\partial^2}{\partial X^2}L_t\big|\mH_t \right]\right)^{-1}=\sigma^2 \mU_t^{-1},
\ee
where $\Exp\left[ -\frac{\partial^2}{\partial X^2}L_t\big|\mH_t \right]$ is the Fisher information matrix and $\mU_t \triangleq \mH_t^T\mH_t$ is a nonsingular matrix.
Since $\Exp[Y_t|\mH_t]=\mH_t X$ and $\Cov(Y_t|\mH_t)=\sigma^2\mI$, from \eqref{eq:LS1} we have $\Exp[\hat{\X}_t|\mH_t]=X$ and $\Cov(\hat{\X}_t|\mH_t)=\sigma^2\mU_t^{-1}$, thus from \eqref{eq:CRLB} $\Cov(\hat{\X}_t|\mH_t)=\text{CRLB}_t$. Note that the maximum likelihood (ML) estimator, that maximizes \eqref{eq:log-like}, coincides with the LS estimator in \eqref{eq:LS1}.

In general, the LS estimator is the best linear unbiased estimator (BLUE). In other words, any linear unbiased estimator of the form $\vec{A}_t Y_t$ with $\vec{A}_t\in\bR^{n\times t}$, where $\Exp[\vec{A}_t Y_t|\mH_t]=X$, has a covariance no smaller than that of the LS estimator in \eqref{eq:LS1}, i.e., $\Cov(\vec{A}_t Y_t|\mH_t)\geq\sigma^2\mU_t^{-1}$ in the positive semidefinite sense. To see this result we write $\vec{A}_t= (\mH_t^T\mH_t)^{-1} \mH_t^T+\vec{B}_t$ for some $\vec{B}_t\in\bR^{n\times t}$, and then $\Cov(\vec{A}_t Y_t|\mH_t)=\sigma^2\mU_t^{-1}+\sigma^2\vec{B}_t\vec{B}_t^T$, where $\vec{B}_t\vec{B}_t^T$ is a positive semidefinite matrix.

The recursive least squares (RLS) algorithm enables us to compute $\hat{\X}_t$ in a recursive way as follows
\begin{align}
\label{eq:RLS}
	\begin{split}
	\hat{\X}_t &= \hat{\X}_{t-1}+K_t (y_t-H_t^T\hat{\X}_{t-1}) \\
	\text{where}~~ K_t &= \frac{\mP_{t-1} H_t}{1+H_t^T\mP_{t-1}H_t}~~\text{and}~~
	\mP_t = \mP_{t-1} - K_t H_t^T \mP_{t-1},
	\end{split}
\end{align}
where $K_t\in\bR^n$ is a gain vector and $\mP_t=\mU_t^{-1}$. While applying RLS we first initialize $\hat{\X}_0=0$ and $\mP_0=\delta^{-1}\mI$, where $0$ represents a zero vector and $\delta$ is a small number, and then at each time $t$ compute $K_t$, $\hat{\X}_t$ and $\mP_t$ as in \eqref{eq:RLS}.

\section{Optimum Sequential Estimation}
\label{sec:opt}

In this section we aim to find the optimal pair $(\cT,\hat{X}_{\cT})$ of stopping time and estimator corresponding to the optimal sequential estimator. The stopping time for a sequential estimator is determined according to a target estimation accuracy. In general, the average stopping time is minimized subject to a constraint on the estimation accuracy, which is a function of the estimator covariance, i.e., 
\be
\label{eq:opt_uncond}
	\min_{\cT,\hat{X}_{\cT}}  \Exp[\cT] ~~\text{s.t.}~~ f\left(\Cov(\hat{X}_{\cT})\right)\leq C,
\ee
where $f(\cdot)$ is a function from $\bR^{n\times n}$ to $\bR$ and $C\in\bR$ is the target accuracy level.

The accuracy function $f$ should be a monotonic function of the covariance matrix $\Cov(\hat{X}_{\cT})$, which is positive semi-definite, in order to make consistent accuracy assessments, e.g., $f(\Cov(\hat{X}_{\cT}))>f(\Cov(\hat{X}_{\cS}))$ for $\cT<\cS$ since $\Cov(\hat{X}_{\cT})\succ\Cov(\hat{X}_{\cS})$ in the positive definite sense. Two popular and easy-to-compute choices are the trace $\Tr(\cdot)$, which corresponds to the mean squared error (MSE), and the Frobenius norm $\|\cdot\|_F$. 
Before handling the problem in \eqref{eq:opt_uncond}, let us explain why we are interested in restricted stopping times that are adapted to a subset of observation history. 

Denote $\{\cF_t\}$ as the filtration that corresponds to the samples $\{y_1,\ldots,y_t\}$ where $\cF_t=\sigma\{y_1,\ldots,y_t\}$ is the $\sigma$-algebra generated by the samples observed up to time $t$, i.e., the accumulated history related to the observed samples, and $\cF_0$ is the trivial $\sigma$-algebra. Similarly we define the filtration $\{\cH_t\}$ where $\cH_t=\sigma\{H_1,\ldots,H_t\}$ and $\cH_0$ is again the trivial $\sigma$-algebra. It is known that, in general, with discrete-time observations and an unrestricted stopping time, that is $\{\cF_t \cup \cH_t\}$-adapted, the sequential CRLB is not attainable under any noise distribution except for the Bernoulli noise \cite{Ghosh87}. On the other hand, in the case of {\em continuous-time observations with continuous paths}, the sequential CRLB is \textcolor{blue}{attained by the LS estimator with} an $\{\cH_t\}$-adapted stopping time, that depends only on $\mH_{\cT}$ \cite{Fellouris12}. Moreover, in the following lemma we show that, with discrete-time observations, the LS estimator attains the conditional sequential CRLB \textcolor{blue}{for the $\{\cH_t\}$-adapted stopping times}. 

\begin{lem}
\label{lem:seq_opt}
With a monotonic accuracy function $f$ and an $\{\cH_t\}$-adapted stopping time $\cT$ we can write
\be
\label{eq:seq_opt}
	f\left( \Cov( \hat{X}_{\cT} | \mH_{\cT}) \right) \geq f \left( \sigma^2 \mU_{\cT}^{-1} \right)
\ee
for all unbiased estimators under Gaussian noise, and for all linear unbiased estimators under non-Gaussian noise, and the LS estimator
\be
\label{eq:LS_est}
	\hat{\X}_{\cT}=\mU_{\cT}^{-1} V_{\cT},~V_{\cT}\triangleq\mH_{\cT}^T Y_{\cT},
\ee
satisfies the inequality in \eqref{eq:seq_opt} with equality.
\end{lem}

\begin{IEEEproof}
Since the LS estimator, with $\Cov(\hat{\X}_t|\mH_t)=\sigma^2\mU_t^{-1}$, is the MVUE under Gaussian noise and the BLUE under non-Gaussian noise, we write
\begin{align}
	f\left(\Cov(\hat{X}_{\cT}|\mH_{\cT})\right) &= f\left( \Exp\left[ \sum_{t=1}^{\infty} (\hat{X}_t-X)(\hat{X}_t-X)^T ~\ind{t=\cT}\big|\mH_t\right] \right) \nn\\
    &= f\left( \sum_{t=1}^{\infty} \Exp\left[ (\hat{X}_t-X)(\hat{X}_t-X)^T \big|\mH_t\right]~\ind{t=\cT} \right) \label{eq:seq_CRLB_0}\\
	&\geq f\left( \sum_{t=1}^{\infty} \sigma^2\mU_t^{-1} ~\ind{t=\cT} \right) \label{eq:seq_CRLB}\\
	&= f\left(\sigma^2\mU_{\cT}^{-1}\right),
\end{align}
for all unbiased estimators under Gaussian noise and for all linear unbiased estimators under non-Gaussian noise. The indicator function $\ind{A}=1$ if $A$ is true, and $0$ otherwise.
We used the facts that the event $\{\cT=t\}$ is $\cH_t$-measurable and $\Exp[(\hat{X}_t-X)(\hat{X}_t-X)^T|\mH_t]=\Cov(\hat{X}_t|\mH_t) \geq \sigma^2\mU_t^{-1}$ to write \eqref{eq:seq_CRLB_0} and \eqref{eq:seq_CRLB}, respectively.
\end{IEEEproof}

\subsection{The Optimum Sequential Estimator}

We are interested in $\{\cH_t\}$-adapted stopping times to use the optimality property of the LS estimator in the sequential sense, shown in Lemma \ref{lem:seq_opt}. 
In this case we assume $\{H_t\}$ is i.i.d..
From the constrained optimization problem in \eqref{eq:opt_uncond}, using a Lagrange multiplier $\lambda$ we obtain the following unconstrained optimization problem,
\be
\label{eq:lagrange1}
    \min_{\cT,\hat{X}_{\cT}} \Exp[\cT] + \lambda f\left(\Cov(\hat{X}_{\cT})\right).
\ee
For simplicity assume a linear accuracy function $f$ so that $f(\Exp[\cdot])=\Exp[f(\cdot)]$, e.g., the trace function $\Tr(\cdot)$. Then, our constraint function becomes the sum of the individual variances, i.e., $\Tr\left(\Cov(\hat{X}_{\cT})\right)=\sum_{i=1}^n \Var(\hat{x}_{\cT}^i)$. Since $\Tr\left(\Cov(\hat{X}_{\cT})\right)=\Tr\left(\Exp\left[\Cov(\hat{X}_{\cT}|\mH_{\cT})\right]\right) = \Exp\left[\Tr\left(\Cov(\hat{X}_{\cT}|\mH_{\cT})\right)\right]$, we rewrite \eqref{eq:lagrange1} as
\be
\label{eq:lagrange2}
    \min_{\cT,\hat{X}_{\cT}} \Exp\left[\cT + \lambda \Tr\left(\Cov(\hat{X}_{\cT}|\mH_{\cT})\right) \right],
\ee
where the expectation is with respect to $\mH_{\cT}$.
From Lemma \ref{lem:seq_opt}, we see that $\Tr\left(\Cov(\hat{X}_{\cT}|\mH_{\cT})\right)$ is minimized by the LS estimator, and so is the objective value in \eqref{eq:lagrange2}.
Hence, $\hat{\X}_{\cT}$ given in \eqref{eq:LS_est} [cf. \eqref{eq:RLS} for recursive computation] is the optimum estimator for the problem in \eqref{eq:opt_uncond}.

Since $\Tr\left(\Cov(\hat{\X}_{\cT}|\mH_{\cT})\right)=\Tr\left( \sigma^2 \mU_{\cT}^{-1} \right)$, to find the optimal stopping time we need to solve the following optimization problem,
\be
\label{eq:lagrange3}
    \min_{\cT} \Exp\left[\cT + \lambda \Tr\left( \sigma^2 \mU_{\cT}^{-1} \right) \right],
\ee
which can be solved by using the \emph{optimal stopping theory}. Writing \eqref{eq:lagrange3} in the following alternative form
\be
\label{eq:dyn1}
    \min_{\cT} \Exp\left[\sum_{t=0}^{\cT-1} 1 + \lambda \Tr\left( \sigma^2 \mU_{\cT}^{-1} \right) \right],
\ee
we see that the term $\sum_{t=0}^{\cT-1} 1$ accounts for the cost of not stopping until time $\cT$ and the term $\lambda \Tr\left( \sigma^2 \mU_{\cT}^{-1} \right)$ represents the cost of stopping at time $\cT$. Note that $\mU_t=\mU_{t-1}+H_t H_t^T$ and given $\mU_{t-1}$ the current state $\mU_t$ is (conditionally) independent of all previous states, hence $\{\mU_t\}$ is a Markov process. That is, in \eqref{eq:dyn1}, the optimal stopping time for a Markov process is sought, which can be found by solving the following Bellman equation
\be
\label{eq:Bellman}
    \cV(\mU) = \min\big\{ \underbrace{\lambda \Tr\left( \sigma^2 \mU^{-1} \right)}_{F(\mU)} , \underbrace{1+\Exp[\cV(\mU+H_1H_1^T)|\mU]}_{G(\mU)} \big\},
\ee
where the expectation is with respect to $H_1$ and $\cV$ is the optimal cost function. The optimal cost function is obtained by iterating a sequence of functions $\{\cV_m\}$ where $\cV(\mU)=\lim_{m\to\infty} \cV_m(\mU)$ and
\be
\label{eq:Bellman_iter}
    \cV_m(\mU) = \min\big\{ \lambda \Tr\left( \sigma^2 \mU^{-1} \right) , 1+\Exp[\cV_{m-1}(\mU+H_1H_1^T)|\mU] \big\}. \nn
\ee

In the above optimal stopping theory, dynamic programming is used. Specifically, the original complex optimization problem in \eqref{eq:lagrange3} is divided into simpler subproblems given by \eqref{eq:Bellman}. At each time $t$ we are faced with a subproblem consisting of a stopping cost $F(\mU_t)=\lambda \Tr\left( \sigma^2 \mU_t^{-1} \right)$ and an expected sampling cost $G(\mU_t)=1+\Exp[\cV(\mU_{t+1})|\mU_t]$ to proceed to time $t+1$. Since $\{\mU_t\}$ is a Markov process, and $\{H_t\}$ is i.i.d., \eqref{eq:Bellman} is a general equation holding for all $t$, and thus we drop the time subscript for simplicity.
The optimal cost function $\cV(\mU_t)$, selecting the action with minimum cost (i.e., either continue or stop), determines the optimal policy to follow at each time $t$. That is, we stop the first time the stopping cost is smaller than the average cost of sampling, i.e.,
\be
    \cT = \min\{t\in\bN: \cV(\mU_t)=F(\mU_t)\}. \nn
\ee
We obviously need to analyze the structure of $\cV(\mU_t)$, i.e., the cost functions $F(\mU_t)$ and $G(\mU_t)$, to find the optimal stopping time $\cT$. We refer to \cite{Shiryaev08} for more information on optimal stopping theory.

Note that $\cV$, being a function of the symmetric matrix $\mU=[u_{ij}]\in\bR^{n\times n}$, is a function of $\frac{n^2+n}{2}$ variables $\{u_{ij}:i\leq j\}$. Analyzing a multi-dimensional optimal cost function proves intractable, hence we will first analyze the special case of scalar parameter estimation and then provide some numerical results for the two-dimensional vector case, demonstrating how intractable the higher dimensional problems are.

\subsubsection{Scalar case}

For the scalar case, from \eqref{eq:Bellman} we have the following one-dimensional optimal cost function,
\be
\label{eq:Bell_sca}
    \cV(u) = \min\left\{ \frac{\lambda \sigma^2}{u} , 1+\Exp[\cV(u+h_1^2)|u] \right\},
\ee
where the expectation is with respect to the scalar coefficient $h_1$. Specifically, at time $t$ the optimal cost function is written as $\cV(u_t) = \min\left\{ \frac{\lambda \sigma^2}{u_t} , 1+\Exp[\cV(u_{t+1})] \right\}$, where $u_{t+1}=u_t+h_{t+1}^2$.
Writing $\cV$ as a function of $z_t\triangleq 1/u_t$ we have $\cV(z_t) = \min\left\{ \lambda \sigma^2 z_t , 1+\Exp[\cV(z_{t+1})] \right\}$, where $z_{t+1}=\frac{z_t}{1+z_t h_{t+1}^2}$, and thus in general
\be
\label{eq:Bell_sca}
    \cV(z) = \min\Bigg\{ \underbrace{\lambda \sigma^2 z}_{F(z)} , \underbrace{1+\Exp\left[\cV\left(\frac{z}{1+z h_1^2}\right)|z\right]}_{G(z)} \Bigg\}.
\ee
We need to analyze the cost functions $F(z)=\lambda \sigma^2 z$ and $G(z)=1+\Exp\left[\cV\left(\frac{z}{1+z h_1^2}\right)|z\right]$. The former is a line, whereas the latter is, in general, a nonlinear function of $z$.
We have the following lemma regarding the structure of $\cV(z)$ and $G(z)$. Its proof is given in the Appendix.

\begin{lem}
\label{lem:unc_sca}
    The optimal cost $\cV$ and the expected sampling cost $G$, given in \eqref{eq:Bell_sca}, are non-decreasing, concave and bounded functions of $z$.
\end{lem}

Following Lemma \ref{lem:unc_sca} the theorem below presents the stopping time for the scalar case of the problem in \eqref{eq:opt_uncond}.

\begin{thm}
\label{thm:unc_sca}
    The optimal stopping time for the scalar case of the problem in \eqref{eq:opt_uncond} with $\Tr(\cdot)$ as the accuracy function is given by
    \be
    \label{eq:st_unc}
    	\cT=\min\left\{ t\in\bN: u_t \geq \frac{1}{C''} \right\},
    \ee
    where $C''$ is selected so that $\Exp\left[ \frac{\sigma^2}{u_{\cT}} \right] = C$, i.e., the variance of the estimator exactly hits the target accuracy level $C$, (see Algorithm \ref{alg:unc_sim_sca}).
\end{thm}

\begin{figure}
\centering
\includegraphics[scale=0.45]{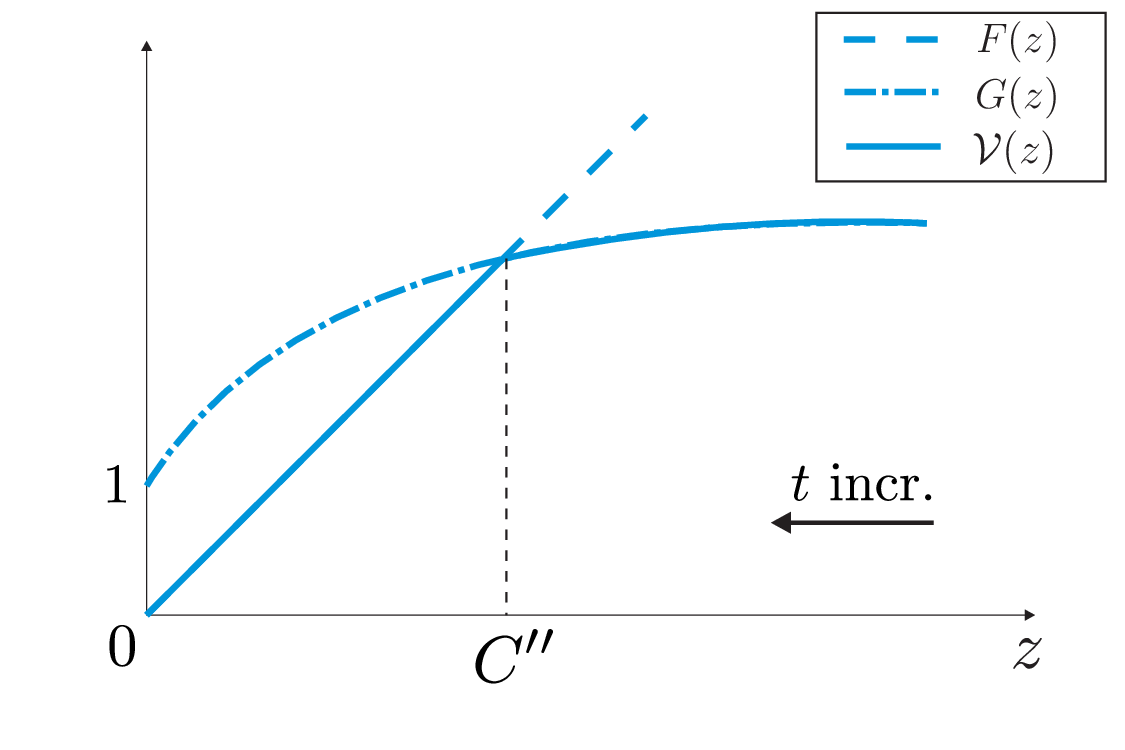}
\caption{The structures of the optimal cost function $\cV(z)$ and the cost functions $F(z)$ and $G(z)$.}
\label{fig:V(z)}
\end{figure}

\begin{IEEEproof}
    The cost functions $F(z)$ and $G(z)$ are continuous functions as $F$ is linear and $G$ is concave. From \eqref{eq:Bell_sca} we have $\cV(0)=\min\{0,1+\cV(0)\}=0$, hence $G(0)=1+\cV(0)=1$.
    Then, using Lemma \ref{lem:unc_sca} we illustrate $F(z)$ and $G(z)$ in Fig. \ref{fig:V(z)}.
    The optimal cost function $\cV(z)$, being the minimum of $F$ and $G$ [cf. \eqref{eq:Bell_sca}], is also shown in Fig. \ref{fig:V(z)}.
    Note that as $t$ increases $z$ tends from infinity to zero.
    Hence, we continue until the stopping cost $F(z_t)$ is lower than the expected sampling cost $G(z_t)$, i.e., until $z_t \leq C''$.
    The threshold $C''(\lambda)=\{z: F(\lambda,z)=G(z)\}$ is determined by the Lagrange multiplier $\lambda$, which is selected to satisfy the constraint $\Var(\hat{x}_{\cT})=\Exp\left[ \frac{\sigma^2}{u_{\cT}} \right] = C$ [cf. \eqref{eq:lagrange1}]. In Algorithm \ref{alg:unc_sim_sca}, we show how to determine the threshold $C''$.
\end{IEEEproof}

We see from Theorem \ref{thm:unc_sca} that the optimum stopping time in the scalar case is given by a threshold rule on the Fisher information. 

\begin{algorithm}[tb]\small
\caption{The procedure to compute the threshold $C''$ for given $C$}
\label{alg:unc_sim_sca}
\baselineskip=0.5cm
  \begin{algorithmic}[1]
    \STATE Select $C''$
    \STATE Estimate $\cC=\Exp\left[ \frac{\sigma^2}{u_{\cT}} \right]$ through simulations, where $u_t=\sum_{p=1}^t h_{p}^2$ and $\cT=\min\left\{ t\in\bN: u_t \geq \frac{1}{C''} \right\}$
    \IF {$\cC=C$}
        \STATE return $C''$
    \ELSE
        \IF {$\cC>C$}
            \STATE Decrease $C''$
        \ELSE
            \STATE Increase $C''$
        \ENDIF
        \STATE Go to line 2
    \ENDIF
  \end{algorithmic}
\end{algorithm}

\subsubsection{Two-dimensional case}

We will next show that the multi-dimensional cases are intractable by providing some numerical results for the two-dimensional case. In the two-dimensional case, we have
\begin{align}
	\Tr\left( \sigma^2 \mU^{-1} \right) &= \sigma^2 \frac{u_{11}+u_{22}}{u_{11}u_{22}-u_{12}^2} \nn\\
	\text{where}~~\mU &= \MB{cc} u_{11} & u_{12} \\ u_{12} & u_{22} \ME ~~,~~ H_1=\MB{c} h_{1,1} \\ h_{1,2} \ME. \nn
\end{align}
Hence, from \eqref{eq:Bellman} the optimal cost function is written as
\be
    \cV(u_{11},u_{12},u_{22}) = \min\left\{ \lambda \sigma^2 \frac{u_{11}+u_{22}}{u_{11}u_{22}-u_{12}^2}, 1+\Exp\left[\cV(u_{11}+h_{1,1}^2,u_{12}+h_{1,1}h_{1,2},u_{22}+h_{1,2}^2)|\mU\right] \right\},
\ee
where the expectation is with respect to $h_{1,1}$ and $h_{1,2}$.
Changing variables we can write $\cV$ as a function of $z_{11}\triangleq1/u_{11}$, $z_{22}\triangleq1/u_{22}$ and $\rho\triangleq u_{12}/\sqrt{u_{11}u_{22}}$,
\begin{multline}
\label{eq:2dim_val}
    \cV(z_{11},z_{22},\rho) = \\
    \min\Bigg\{ \underbrace{\lambda \sigma^2 \frac{z_{11}+z_{22}}{1-\rho^2}}_{F(z_{11},z_{22},\rho)},
    \underbrace{1+\Exp\Bigg[ \cV\bigg( \frac{z_{11}}{1+z_{11}h_{1,1}^2}, \frac{z_{22}}{1+z_{22}h_{1,2}^2}, \frac{\rho+h_{1,1}h_{1,2}\sqrt{z_{11}z_{22}}}{\sqrt{(1+z_{11}h_{1,1}^2)(1+z_{22}h_{1,2}^2)}} \bigg)\Big|z_{11},z_{22},\rho \Bigg]}_{G(z_{11},z_{22},\rho)} \Bigg\},
\end{multline}
which can be iteratively computed as follows
\begin{multline}
\label{eq:2dim_val_it}
    \cV_m(z_{11},z_{22},\rho) =\\
    \min\Bigg\{ \lambda \sigma^2 \frac{z_{11}+z_{22}}{1-\rho^2},
    1+\Exp\Bigg[ \cV_{m-1}\bigg( \frac{z_{11}}{1+z_{11}h_{1,1}^2}, \frac{z_{22}}{1+z_{22}h_{1,2}^2}, \frac{\rho+h_{1,1}h_{1,2}\sqrt{z_{11}z_{22}}}{\sqrt{(1+z_{11}h_{1,1}^2)(1+z_{22}h_{1,2}^2)}} \bigg)\Big|z_{11},z_{22},\rho \Bigg] \Bigg\},
\end{multline}
where $\lim_{m\to\infty}\cV_m=\cV$.

Note that $\rho$ is the correlation coefficient, hence we have $\rho\in[-1,1]$.
Following the procedure in Algorithm \ref{alg:unc_sim_surf} we numerically compute $\cV$ from \eqref{eq:2dim_val_it} and find the boundary surface
\be
    \mathscr{S}(\lambda)=\{(z_{11},z_{22},\rho): F(\lambda,z_{11},z_{22},\rho)=G(z_{11},z_{22},\rho)\}, \nn
\ee
that defines the stopping rule.
\begin{algorithm}[thb]\small
\caption{The procedure to compute the boundary surface $\mathscr{S}$ for given $\lambda$}
\label{alg:unc_sim_surf}
\baselineskip=0.5cm
  \begin{algorithmic}[1]
    \STATE Set dz, Rz, dr, Nh; ~~$\text{Nz}=\frac{\text{Rz}}{\text{dz}}+1$; ~~$\text{Nr}=\frac{2}{\text{dr}}+1$
    \STATE $z_1=[0:\text{dz}:\text{Rz}];~~z_2=z_1;~~\rho=[-1:\text{dr}:1]$ ~~\COMMENT{all row vectors}
    \STATE $Z_1=\vec{1}_{\text{Nz}}z_1;~~Z_2=Z_1^T$ ~~\COMMENT{$\vec{1}_{\text{Nz}}$: column vector of ones in $\bR^{\text{Nz}}$}
    \FOR {$i=1:\text{Nr}$}
        \STATE $F(:,:,i)=\lambda\frac{Z_1+Z_2}{1-\rho(i)^2}$ ~~\COMMENT{stopping cost over the 3D grid}
    \ENDFOR
    \STATE $\cV=\min(F,1)$ ~~\COMMENT{start with $\cV_0=0$}
    \STATE $\text{dif}=\infty$;~~$\text{Fr}=\|\cV\|_F$
    \WHILE {$\text{dif}>\delta ~\text{Fr}$ ~~\COMMENT{$\delta$: a small threshold}}
        \FOR {$i=1:\text{Nz}^2$}
            \STATE $z_{11}=Z_1(i)$;~~$z_{22}=Z_2(i)$ ~~\COMMENT{linear indexing in matrices}
            \FOR {$j=1:\text{Nr}$}		                
	       	\STATE Generate $h_1^{\text{Nh}\times 1}$ and $h_2^{\text{Nh}\times 1}$ ~~\COMMENT{e.g., according to $\cN(0,1)$}
                    \STATE $Z_{11}'=z_{11}./(1+z_{11}h_1.^2)$; $Z_{22}'=z_{22}./(1+z_{22}h_2.^2)$ ~~\COMMENT{dot denotes elementwise operation}
                    \STATE $\rho'=[\rho(j)+h_1.*h_2 \sqrt{z_{11} z_{22}}]./\sqrt{(1+z_{11} h_1.^2)(1+z_{22} h_2.^2)}$ ~\COMMENT{vector}
                    \STATE $I_1=Z_{11}'/\text{dz}+1$;~~$I_2=Z_{22}'/\text{dz}+1$;~~$I_3=(\rho'+1)/\text{dr}+1$ ~~\COMMENT{fractional indices}
                    \STATE $J^{8\times \text{Nh}}=$ linear indices of 8 neighbor points using $\lfloor I_n \rfloor$, $\lceil I_n \rceil$, $n=1,2,3$
                    \STATE $D_n=\lceil I_n \rceil-I_n$;~~$\overline{D}_n=1-D_n,~n=1,2,3$ ~~\COMMENT{distances to neighbor indices}
                    \STATE $W^{8\times \text{Nh}}=$ weights for neighbors as 8 multiplicative combinations of $D_n$, $\overline{D}_n$, $n=1,2,3$
                    \STATE $V^{\text{Nh}\times 1}=\text{diag}(W^T \cV(J))$ ~~\COMMENT{average the neighbor $\cV$ values}
                    \STATE $G=\text{sum}(V)/\text{Nh}$ ~~\COMMENT{continuing cost}
                \STATE $\ell=i+(j-1)\text{Nz}^2$ ~~\COMMENT{linear index of the point the 3D grid}
                \STATE $\cV'(\ell)=\min(F(\ell),1+G)$ ~~\COMMENT{new optimal cost function}
            \ENDFOR
        \ENDFOR
        \STATE $\text{dif}=\|\cV'-\cV\|_F$;~~$\text{Fr}=\|\cV\|_F$
        \STATE $\cV=\cV'$ ~~\COMMENT{update the optimal cost function}
    \ENDWHILE
    \STATE Find the points where transition occurs between regions $\cV=F$ and $\cV\not=F$, i.e., $\mathscr{S}$.
  \end{algorithmic}
\end{algorithm}
In Algorithm \ref{alg:unc_sim_surf}, firstly the three-dimensional grid $(n_1 dz,n_2 dz,n_3 dr),~\\n_1,n_2=0,\ldots,\frac{\text{Rz}}{\text{dz}},~n_3=-\frac{1}{\text{dr}},\ldots,\frac{1}{\text{dr}}$ is constructed. Then, in lines 4-6 the stopping cost $F$ [cf. \eqref{eq:2dim_val}] and in line 7 the first iteration of the optimal cost function $\cV_1$ with $\cV_0=0$ are computed over the grid. In lines 9-28, the optimal cost function $\cV$ is computed for each point in the grid by iterating $\cV_m$ [cf. \eqref{eq:2dim_val_it}] until no significant change occurs between $\cV_m$ and $\cV_{m+1}$. In each iteration, in lines 13-21, the expectation in \eqref{eq:2dim_val_it} with respect to $h_{1,1}$ and $h_{1,2}$ is computed through Monte Carlo calculations. While computing the expectation, since the updated (future) $(z_{11},z_{22},\rho)$ values, i.e, the arguments of $\cV_{m-1}$ in \eqref{eq:2dim_val_it}, in general may not correspond to a grid point, we average the $\cV_{m-1}$ values of eight neighboring grid points with appropriate weights in lines 17-20 to obtain the desired $\cV_{m-1}$ value.

\begin{figure}
\centering
\includegraphics[scale=0.5]{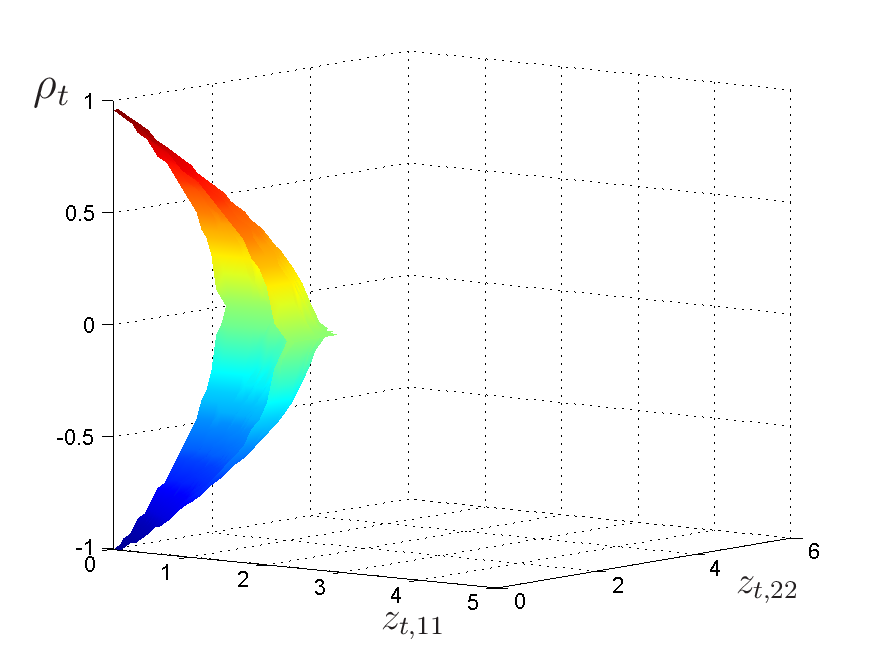}
\caption{The surface that defines the stopping rule for $\lambda=1$, $\sigma^2=1$ and $h_{1,1},h_{1,2}\sim\cN(0,1)$ in the two-dimensional case.}
\label{fig:3d}
\end{figure}

\begin{figure}
\centering
\includegraphics[scale=0.55]{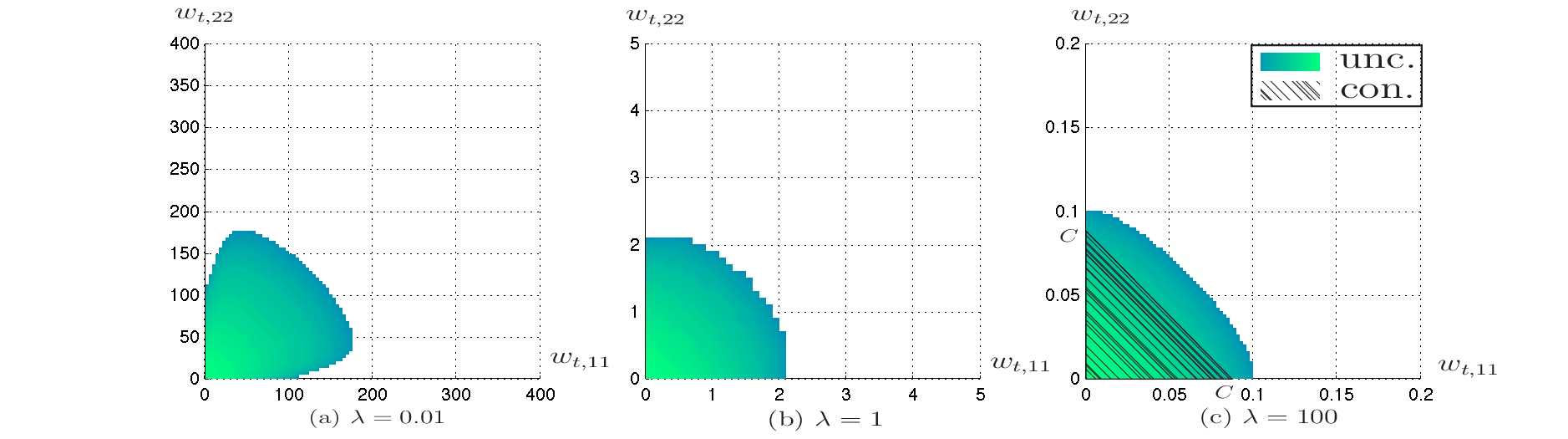}
\caption{The stopping regions for $\rho_t=0$, $\sigma^2=1$ and $h_{t,1},h_{t,2}\sim\cN(0,1),\forall t$ with (a) $\lambda=0.01$, (b) $\lambda=1$, (c) $\lambda=100$. That of the conditional problem (see Section \ref{sec:cond}) is also shown in (c).}
\label{fig:2d}
\end{figure}

The results for $\lambda\in\{0.01,1,100\}$, $\sigma^2=1$ and $h_{1,1},h_{1,2}\sim\cN(0,1)$ are shown in Fig. \ref{fig:3d} and Fig. \ref{fig:2d}.
For $\lambda=1$, the dome-shaped surface in Fig. \ref{fig:3d} separates the stopping region from the continuing region. Outside the ``dome" $\cV=G$, hence we continue. As time progresses $z_{t,11}$ and $z_{t,22}$ decrease, so we move towards the ``dome". And whenever we are inside the ``dome", we stop, i.e., $\cV=F$.
We obtain similar dome-shaped surfaces for different $\lambda$ values. However, the cross-sections of the ``domes" at specific $\rho_t$ values differ significantly.
In particular, we investigate the case of $\rho_t=0$, where the scaling coefficients $h_{t,1}$ and $h_{t,2}$ are uncorrelated. For small values of $\lambda$, e.g., $\lambda=0.01$, the boundary that separates the stopping and the continuing regions is highly nonlinear as shown in Fig. \ref{fig:2d}(a). In Fig. \ref{fig:2d}(b) and \ref{fig:2d}(c), it is seen that the boundary tends to become more and more linear as $\lambda$ increases.

Now let us explain the meaning of the $\lambda$ value. Firstly, note from \eqref{eq:2dim_val} that $F$ and $G$ are functions of $z_{11}$, $z_{22}$ for fixed $\rho$, and the boundary is the solution to $F(\lambda,z_{11},z_{22})=G(z_{11},z_{22})$. When $\lambda$ is small, the region where $F<G$, i.e., the stopping region, is large, hence we stop early as shown in Fig. \ref{fig:2d}(a) \footnote{Note that the axis scales in Fig. \ref{fig:2d}(a) are on the order of hundreds and $z_{t,11}$, $z_{t,22}$ decrease as $t$ increases.}. Conversely, for large $\lambda$ the stopping region is small, hence the stopping time is large [cf. Fig. \ref{fig:2d}(c)].
\begin{algorithm}[tb]\small
\caption{The procedure to compute the boundary surface $\mathscr{S}$}
\label{alg:unc_sim}
\baselineskip=0.5cm
  \begin{algorithmic}[1]
    \STATE Select $\lambda$
    \STATE Compute $\mathscr{S}(\lambda)$ as in Algorithm \ref{alg:unc_sim_surf}
    \STATE Estimate $\cC=\Exp\left[ \sigma^2 \frac{z_{\cT,11}+z_{\cT,22}}{1-\rho_\cT^2} \right]$ through simulations, where $z_{t,11}=1/u_{t,11}$, $z_{t,22}=1/u_{t,22}$, $\rho_t=u_{t,12}/\sqrt{u_{t,11}u_{t,22}}$ and $\cT=\min\{t\in\bN:(z_{t,11},z_{t,22},\rho_t) ~\text{is between $\mathscr{S}$ and the origin}\}$
    \IF {$\cC=C$}
        \STATE return $\mathscr{S}$
    \ELSE
        \IF {$\cC>C$}
            \STATE Increase $\lambda$
        \ELSE
            \STATE Decrease $\lambda$
        \ENDIF
        \STATE Go to line 2
    \ENDIF
  \end{algorithmic}
\end{algorithm}
In fact, the Lagrange multiplier $\lambda$ is selected through simulations following the procedure in Algorithm \ref{alg:unc_sim} so that the constraint $\Tr\Big(\Exp\big[\sigma^2 \mU_{\cT}^{-1}\big]\Big)=\Exp\left[ \sigma^2 \frac{z_{\cT,11}+z_{\cT,22}}{1-\rho_\cT^2} \right]= C$ is satisfied. Note that line 2 of Algorithm \ref{alg:unc_sim} uses Algorithm \ref{alg:unc_sim_surf} to compute the boundary surface $\mathscr{S}$.

\vspace{3mm}
\underline{\bf\em Remarks:}
In general, we need to numerically compute the stopping rule offline, i.e., the hypersurface that separates the stopping and the continuing regions, for a given target accuracy level $C$. This becomes a quite intractable task as the dimension $n$ of the vector to be estimated increases as we find the separating hypersurface in a $\frac{n^2+n}{2}$-dimensional space. Recall from \eqref{eq:Bellman} that the optimal cost function $\cV$ is a function of the matrix $\mU$, which has $\frac{n^2+n}{2}$ distinct entries. On the other hand, conditioning the problem formulation in \eqref{eq:opt_uncond} on the observed regressors $\{H_t\}$, we next show that, for any $n$, the optimum stopping rule takes a simple one-dimensional form. We can much more easily decentralize such a tractable optimum solution offered by the conditional formulation than the one given by the cumbersome procedure in Algorithm \ref{alg:unc_sim_surf}.

\subsection{The Optimum Conditional Sequential Estimator}
\label{sec:cond}

In the presence of an ancillary statistic whose distribution does not depend on the parameters to be estimated, such as the regressor matrix $\mH_t$, the conditional covariance $\Cov(\hat{X}_t|\mH_t)$ can be used to assess the accuracy of the estimator more precisely than the (unconditional) covariance, which is in fact the mean of the former, i.e., $\Cov(\hat{X}_{\cT})=\Exp[\Cov(\hat{X}_t|\mH_t)]$, \cite{Efron78,Grambsch83}. Motivated by this fact we propose to reformulate the problem in \eqref{eq:opt_uncond} conditioned on $\mH_t$, that is,
\be
\label{eq:opt_cond}
  \min_{\cT,\hat{X}_{\cT}}  \Exp[\cT] ~~\text{s.t.}~~ f\left(\Cov(\hat{X}_{\cT}|\mH_{\cT})\right)\leq C.
\ee

Note that the constraint in \eqref{eq:opt_cond} is stricter than the one in \eqref{eq:opt_uncond} since it requires that $\hat{X}_{\cT}$ satisfies the target accuracy level for each realization of $\mH_{\cT}$, whereas in \eqref{eq:opt_uncond} it is sufficient that $\hat{X}_{\cT}$ satisfies the target accuracy level on average. In other words, in \eqref{eq:opt_uncond}, even if $f\left(\Cov(\hat{X}_{\cT}|\mH_{\cT})\right)> C$ for some realizations of $\mH_{\cT}$, we can still satisfy $f\left(\Cov(\hat{X}_{\cT})\right)\leq C$. In fact, we can always have $f\left(\Cov(\hat{X}_{\cT})\right)=C$ by using a probabilistic stopping rule such that we sometimes stop above $C$, i.e., $f\left(\Cov(\hat{X}_{\cT}|\mH_{\cT})\right)> C$, and the rest of the time at or below $C$, i.e., $f\left(\Cov(\hat{X}_{\cT}|\mH_{\cT})\right)\leq C$. On the other hand, in \eqref{eq:opt_cond} we always have $f\left(\Cov(\hat{X}_{\cT}|\mH_{\cT})\right)\leq C$, and
moreover since we observe discrete-time samples, in general we have $f\left(\Cov(\hat{X}_{\cT}|\mH_{\cT})\right)<C$ for each realization of $\mH_{\cT}$.
Hence, the optimal objective value $\Exp[\cT]$ in \eqref{eq:opt_uncond} will, in general, be smaller than that in \eqref{eq:opt_cond}. Note that on the other hand, if we observed continuous-time processes with continuous paths, then we could always have $f\left(\Cov(\hat{X}_{\cT}|\mH_{\cT})\right)=C$ for each realization of $\mH_{\cT}$, and thus the optimal objective values of \eqref{eq:opt_cond} and \eqref{eq:opt_uncond} would be the same.

Since minimizing $\cT$ also minimizes $\Exp[\cT]$, in \eqref{eq:opt_cond} we want to find the first time that a member of our class of estimators (i.e., unbiased estimators under Gaussian noise and linear unbiased estimators under non-Gaussian noise) satisfies the constraint $f\left(\Cov(\hat{X}_{\cT}|\mH_{\cT})\right)\leq C$, as well as the estimator that attains this earliest stopping time.
From Lemma \ref{lem:seq_opt}, it is seen that the LS estimator, given by \eqref{eq:LS_est}, among its competitors, achieves the best accuracy level $f\left(\sigma^2\mU_{\cT}^{-1}\right)$ at any stopping time $\cT$. Hence, for the conditional problem the optimum sequential estimator is composed of the stopping time
\be
\label{eq:stop_time}
	\cT = \min\{t\in \bN: f\left(\sigma^2\mU_t^{-1}\right)\leq C\},
\ee
and the LS estimator
\be
\label{eq:LS2}
    \hat{\X}_{\cT}=\mU_{\cT}^{-1} V_{\cT},
\ee
which can be computed recursively as in \eqref{eq:RLS}.
The recursive computation of $\mU_t^{-1}=\mP_t$ in the test statistic in \eqref{eq:stop_time} is also given in \eqref{eq:RLS}. Note that for an accuracy function $f$ such that $f(\sigma^2\mU_t^{-1})=\sigma^2f(\mU_t^{-1})$, e.g., $\Tr(\cdot)$ and $\|\cdot\|_F$, we can use the following stopping time,
\be
\label{eq:stop_time1}
	\cT = \min\{t\in \bN: f\left(\mU_t^{-1}\right)\leq C^\prime\},
\ee
where $C^\prime=\frac{C}{\sigma^2}$ is the relative target accuracy with respect to the noise power. Hence, given $C^\prime$ we do not need to know the noise variance $\sigma^2$ to run the test given by \eqref{eq:stop_time1}.
Note that $\mU_t=\mH_t^T \mH_t$ is a non-decreasing positive semi-definite matrix, i.e., $\mU_t\succeq\mU_{t-1},\forall t$, in the positive semi-definite sense. Thus, from the monotonicity of $f$, the test statistic $f\left(\sigma^2\mU_t^{-1}\right)$ is a non-increasing scalar function of time. Specifically, for accuracy functions $\Tr(\cdot)$ and $\|\cdot\|_F$ we can show that if the minimum eigenvalue of $\mU_t$ tends to infinity as $t\to\infty$, then the stopping time is finite, i.e., $\cT<\infty$.

In the conditional problem, for any $n$, we have a simple stopping rule given in \eqref{eq:stop_time1}, which uses the target accuracy level $\frac{C}{\sigma^2}$ as its threshold, hence known beforehand. 
For the special case of scalar parameter estimation, we do not need a function $f$ to assess the accuracy of the estimator since instead of a covariance matrix we now have a variance $\frac{\sigma^2}{u_t}$, where $u_t=\sum_{p=1}^t h_{p}^2$ and $h_t$ is the scaling coefficient in \eqref{eq:sig_mod}. Hence, from \eqref{eq:stop_time1} the stopping time in the scalar case is given by
\be
\label{eq:stop_time_sca}
	\cT = \min\left\{t\in \bN: u_t \geq \frac{1}{C^\prime}\right\},
\ee
where $\frac{u_t}{\sigma^2}$ is the Fisher information at time $t$. That is, we stop the first time the gathered Fisher information exceeds the threshold $1/C'$, which is known.

Note that the optimal stopping time in the scalar case of the unconditional problem, given by \eqref{eq:st_unc}, is of the same form as in \eqref{eq:stop_time_sca}. In both conditional and unconditional problems the LS estimator
\be
    \hat{\x}_{\cT} = \frac{v_{\cT}}{u_{\cT}} \nn
\ee
is the optimal estimator.
The fundamental difference between the optimal stopping times in \eqref{eq:stop_time_sca} and \eqref{eq:st_unc} is that the threshold $C^\prime=\frac{C}{\sigma^2}$ in the conditional problem is known beforehand; whereas the threshold $C''$ in the unconditional problem needs to be determined through offline simulations following the procedure in Algorithm \ref{alg:unc_sim_sca}.
We also observe that $C'\leq C''$, hence the optimal objective value $\Exp[\cT]$ of the unconditional problem is in general smaller than that of the conditional problem as noted earlier in this subsection. This is because the upper bound $\sigma^2 C''$ on the conditional variance $\frac{\sigma^2}{u_{\cT}}$ [cf. \eqref{eq:st_unc}] is also an upper bound for the variance $\Exp\big[\frac{\sigma^2}{u_{\cT}}\big]=C$, and the threshold $C'$ is given by $C'=\frac{C}{\sigma^2}$.

In the two-dimensional case of the conditional problem the optimal stopping time is given by $\cT=\min\left\{ t\in\bN: \frac{z_{t,11}+z_{t,22}}{1-\rho_t^2}\leq \frac{C}{\sigma^2} \right\}$, which is a function of $z_{t,11}+z_{t,22}$ for fixed $\rho_t$. In Fig. \ref{fig:2d}(c), where $\rho_t=0$ and $\sigma^2=1$, the stopping region (resp. average stopping time) of the conditional problem, which is characterized by a line, is shown to be smaller (resp. larger) than that of the unconditional problem due to the same reasoning in the scalar case.

\section{Decentralized Sequential Estimation}
\label{sec:decent}

In this section, we propose a computation- and energy-efficient decentralized estimator based on the optimum conditional sequential estimator and level-triggered sampling. Consider a network of $K$ distributed sensors and a fusion center (FC) which is responsible for determining the stopping time and computing the estimator. In practice, due to the stringent energy constraints, sensors must \emph{infrequently} convey \emph{low-rate} information to the FC, which is the main concern in the design of a decentralized sequential estimator.

As in \eqref{eq:sig_mod} each sensor $k$ observes
\be
\label{eq:sig_mod_dec}
    y_t^k = (H_t^k)^T X + w_t^k, ~~t\in\bN, ~~k=1,\ldots,K,
\ee
as well as the regressor vector $H_t^k=[h_{t,1}^k,\ldots,h_{t,n}^k]^T$ at time $t$, where $\{w_t^k\}_{k,t}$ \footnote{The subscripts $k$ and $t$ in the set notation denote $k=1,\ldots,K$ and $t\in\bN$.} are independent, zero-mean, i.e., $\Exp[w_t^k]=0,~\forall k,t$, and $\Var(w_t^k)=\sigma^2_k,~\forall t$. Then, similar to \eqref{eq:LS1} the weighted least squares (WLS) estimator
$$
\hat{\X}_t = \arg \min_{X} \sum_{k=1}^K \sum_{p=1}^t \frac{\left(y_{p}^k-(H_{p}^k)^T X\right)^2}{\sigma_k^2}
$$
is given by
\be
\label{eq:WLS1_dec}
    \hat{\X}_t = \left(\sum_{k=1}^K \sum_{p=1}^t \frac{H_{p}^k (H_{p}^k)^T}{\sigma_k^2}\right)^{-1} \sum_{k=1}^K \sum_{p=1}^t \frac{H_{p}^k y_{p}^k}{\sigma_k^2} = \bar{\mU}_t^{-1} \bar{V}_t
\ee
where $\bar{\mU}_t^k \triangleq \frac{1}{\sigma_k^2}\sum_{p=1}^t H_{p}^k (H_{p}^k)^T$, $\bar{V}_t^k \triangleq \frac{1}{\sigma_k^2}\sum_{p=1}^t H_{p}^k y_{p}^k$, $\bar{\mU}_t=\sum_{k=1}^K \bar{\mU}_t^k$ and $\bar{V}_t=\sum_{k=1}^K \bar{V}_t^k$.
As before it can be shown that the WLS estimator $\hat{\X}_t$ in \eqref{eq:WLS1_dec} is the BLUE under the general noise distributions. Moreover, in the Gaussian noise case, where $w_t^k\sim\cN(0,\sigma_k^2)~\forall t$ for each $k$, $\hat{\X}_t$ is also the MVUE. 

Following the steps in Section \ref{sec:cond} it is straightforward to show that the optimum sequential estimator for the conditional problem in \eqref{eq:opt_cond} is given by the stopping time 
\be
\label{eq:stop_time_dec}
	\cT = \min\left\{t\in \bN: f\left(\bar{\mU}_t^{-1}\right)\leq C\right\},
\ee
and the WLS estimator $\hat{\X}_{\cT}$ [cf. \eqref{eq:WLS1_dec}]. Note that $(\cT,\hat{\X}_{\cT})$ is achievable only in the centralized case, where all local observations until time $t$, i.e., $\{(y_{p}^k,H_{p}^k)\}_{k,p}$ \footnote{The subscript $p$ in the set notation denotes $p=1,\ldots,t$.}, are available to the FC. Local processes $\{\bar{\mU}_t^k\}_{k,t}$ and $\{\bar{V}_t^k\}_{k,t}$ are used to compute the stopping time and the estimator as in \eqref{eq:stop_time_dec} and \eqref{eq:WLS1_dec}, respectively. On the other hand, in a decentralized system the FC can compute approximations $\widetilde{\mU}_t^k$ and $\widetilde{V}_t^k$, and then use these approximations
to compute the stopping time and estimator as in \eqref{eq:stop_time_dec} and \eqref{eq:WLS1_dec}, respectively.

\subsection{Key Approximations in Decentralized Approach}
\label{sec:simp}

If each sensor $k$ reports $\bar{\mU}_t^k\in\bR^{n\times n}$ and $\bar{V}_t^k\in\bR^{n}$ to the FC in a straightforward way, then $O(n^2)$ terms need to be transmitted, which may not be practical, especially for large $n$, in a decentralized setup. Similarly, in the literature, the distributed implementation of the Kalman filter, which covers RLS as a special case, through its inverse covariance form, namely the information filter, requires the transmission of an $n \times n$ information matrix and an $n \times 1$ information vector e.g., \cite{Vercauteren05}. 

To overcome this problem, considering $\Tr(\cdot)$ as the accuracy function $f$ in \eqref{eq:stop_time_dec}, we propose to transmit only the $n$ diagonal entries of $\bar{\mU}_t^k$ for each $k$, yielding linear complexity $O(n)$.
Using the diagonal entries of $\bar{\mU}_t$ we define the diagonal matrix
\begin{align}
\label{eq:diag}
\begin{split}
\mD_t &\triangleq \text{diag}\left( d_{t,1},\ldots,d_{t,n} \right) \\
\text{where}~~d_{t,i} &= \sum_{k=1}^K \sum_{p=1}^t \frac{(h_{p,i}^k)^2}{\sigma_k^2},~i=1,\ldots,n.
\end{split}
\end{align}
We further define the correlation matrix
\begin{align}
\label{eq:matrix_R}
  \mR &= \MB{cccc} 1 & r_{12} & \cdots & r_{1n} \\
                    r_{12} & 1 & \cdots & r_{2n} \\
                    \vdots & \vdots & \ddots & \vdots \\
                    r_{1n} & r_{2n} & \cdots & 1 \ME, \\
  \text{where}~~ r_{ij} &= \frac{\sum_{k=1}^K \frac{\Exp[h_{t,i}^k h_{t,j}^k]}{\sigma_k^2}} {\sqrt{\sum_{k=1}^K \frac{\Exp[(h_{t,i}^k)^2]}{\sigma_k^2} \sum_{k=1}^K \frac{\Exp[(h_{t,j}^k)^2]}{\sigma_k^2} }}, ~i,j=1,\ldots,n. \nn
\end{align}

\begin{pro}
\label{pro:approx}
For sufficiently large $t$, we can make the following approximations,
\begin{align}
\label{eq:dec_approx1}
\begin{split}
\bar{\mU}_t &\cong \mD_t^{1/2} \mR ~\mD_t^{1/2} \\
\text{and} ~~\Tr\left(\bar{\mU}_t^{-1}\right) &\cong \Tr\left( \mD_t^{-1} \mR^{-1} \right).
\end{split}
\end{align}
\end{pro}

\begin{IEEEproof}
The approximations are motivated from the special case where $\Exp[h_{t,i}^k h_{t,j}^k]=0,~\forall k,~i,j=1,\ldots,n,~i\not=j$. In this case, by the \emph{law of large numbers} for sufficiently large $t$ the off-diagonal elements of $\frac{\bar{\mU}_t}{t}$ vanish, and thus we have $\frac{\bar{\mU}_t}{t} \cong \frac{\mD_t}{t}$ and $\Tr(\bar{\mU}_t^{-1})\cong\Tr(\mD_t^{-1})$.
For the general case where we might have $\Exp[h_{t,i}^k h_{t,j}^k]\not=0$ for some $k$ and $i\not=j$, using the diagonal matrix $\mD_t$ we write
\begin{align}
\label{eq:tr_uncor1}
    \Tr\left(\bar{\mU}_t^{-1}\right) &= \Tr\Bigg( \bigg( \mD_t^{1/2} \underbrace{\mD_t^{-1/2} \bar{\mU}_t \mD_t^{-1/2}}_{\mR_t} \mD_t^{1/2} \bigg)^{-1} \Bigg) \\
    &= \Tr\left( \mD_t^{-1/2} \mR_t^{-1} \mD_t^{-1/2} \right) \nn\\
    \label{eq:tr_uncor2}
    &= \Tr\left( \mD_t^{-1} \mR_t^{-1} \right).
\end{align}
Note that each entry $r_{t,ij}$ of the newly defined matrix $\mR_t$ is a normalized version of the corresponding entry $\bar{u}_{t,ij}$ of $\bar{\mU}_t$. Specifically, $r_{t,ij}=\frac{\bar{u}_{t,ij}}{\sqrt{d_{t,i}d_{t,j}}}=\frac{\bar{u}_{t,ij}}{\sqrt{\bar{u}_{t,ii}\bar{u}_{t,jj}}},~ i,j=1,\ldots,n$, where the last equality follows from the definition of $d_{t,i}$ in \eqref{eq:diag}. Hence, $\mR_t$ has the same structure as in \eqref{eq:matrix_R} with entries
\be
r_{t,ij} = \frac{\sum_{k=1}^K \sum_{p=1}^t \frac{h_{p,i}^k h_{p,j}^k}{\sigma_k^2}} {\sqrt{\sum_{k=1}^K \sum_{p=1}^t \frac{(h_{p,i}^k)^2}{\sigma_k^2} \sum_{k=1}^K \sum_{p=1}^t \frac{(h_{p,j}^k)^2}{\sigma_k^2} }}, ~i,j=1,\ldots,n. \nn
\ee
For sufficiently large $t$, by the law of large numbers
\be
\label{eq:uncor3}
	r_{t,ij} \cong r_{ij} = \frac{\sum_{k=1}^K \frac{\Exp[h_{t,i}^k h_{t,j}^k]}{\sigma_k^2}} {\sqrt{\sum_{k=1}^K \frac{\Exp[(h_{t,i}^k)^2]}{\sigma_k^2} \sum_{k=1}^K \frac{\Exp[(h_{t,j}^k)^2]}{\sigma_k^2} }}
\ee
and $\mR_t \cong \mR$, where $\mR$ is given in \eqref{eq:matrix_R}. Hence, for sufficiently large $t$ we can make the approximations in \eqref{eq:dec_approx1} using \eqref{eq:tr_uncor1} and \eqref{eq:tr_uncor2}.
\end{IEEEproof}

Then, assuming that the FC knows the correlation matrix $\mR$, i.e., $\big\{\Exp[h_{t,i}^k h_{t,j}^k]\big\}_{i,j,k}$
\footnote{The subscripts $i$ and $j$ in the set notation denote $i=1,\ldots,n$ and $j=i,\ldots,n$. In the special case where $\Exp[(h_{t,i}^k)^2]=\Exp[(h_{t,i}^m)^2],~k,m=1,\ldots,K,~i=1,\ldots,n$, the correlation coefficients
\be
	\left\{\xi_{ij}^k=\frac{\Exp[h_{t,i}^k h_{t,j}^k]}{\sqrt{\Exp[(h_{t,i}^k)^2] \Exp[(h_{t,j}^k)^2]}}:~i=1,\ldots,n-1,~j=i+1,\ldots,n \right\}_k \nn
\ee
together with $\big\{\sigma_k^2\big\}$ are sufficient statistics since $r_{ij}=\frac{\sum_{k=1}^K \xi_{ij}^k/\sigma_k^2} {\sum_{k=1}^K 1/\sigma_k^2}$ from \eqref{eq:uncor3}.}
and $\big\{\sigma_k^2\big\}$ [cf. \eqref{eq:matrix_R}], it can compute the approximations in \eqref{eq:dec_approx1} if sensors report their local processes $\left\{\mD_t^k\right\}_{k,t}$ to the FC, where $\mD_t=\sum_{k=1}^K \mD_t^k$. Note that each local process $\left\{\mD_t^k\right\}_t$ is $n$-dimensional, and its entries at time $t$ are given by $\left\{d_{t,i}^k=\sum_{p=1}^t \frac{(h_{p,i}^k)^2}{\sigma_k^2}\right\}_i$ [cf. \eqref{eq:diag}]. Hence, we propose that each sensor $k$ sequentially reports the local processes $\{\mD_t^k\}_t$ and $\{\bar{V}_t^k\}_t$ to the FC, achieving linear complexity $O(n)$. On the other side, the FC, using the information received from sensors, computes the approximations $\{\widetilde{\mD}_t\}$ and $\{\widetilde{V}_t\}$, which are then used to compute the stopping time
\be
\label{eq:stop_time_dec2}
	\widetilde{\cT} = \min\left\{t\in \bN: \Tr\left(\widetilde{\mU}_t^{-1}\right)\leq \widetilde{C}\right\},
\ee
and the estimator
\be
\label{eq:est_dec}
    \widetilde{X}_{\widetilde{\cT}} = \widetilde{\mU}_{\widetilde{\cT}}^{-1} \widetilde{V}_{\widetilde{\cT}}
\ee
similar to \eqref{eq:stop_time_dec} and \eqref{eq:WLS1_dec}, respectively. The approximations $\Tr\left(\widetilde{\mU}_t^{-1}\right)$ in \eqref{eq:stop_time_dec2}  and $\widetilde{\mU}_{\widetilde{\cT}}$ in \eqref{eq:est_dec} are computed using $\widetilde{\mD}_t$ as in \eqref{eq:dec_approx1}. The threshold $\widetilde{C}$ is selected through simulations to satisfy the constraint in \eqref{eq:opt_cond} with equality, i.e., $\Tr\left( \Cov\big( \widetilde{X}_{\widetilde{\cT}}|\mH_{\widetilde{\cT}} \big) \right) = C$.

\subsection{Decentralized Sequential Estimator Based on Level-triggered Sampling}
\label{sec:dest}

Level-triggered sampling provides a very convenient way of information transmission in decentralized systems \cite{Fellouris11,Yilmaz14}. Specifically, decentralized methods based on level-triggered sampling, transmitting low-rate information, enable highly accurate approximations and thus high performance schemes at the FC. They significantly outperform conventional decentralized methods which sample local processes using the traditional uniform sampling and send the quantized versions of samples to the FC \cite{Yilmaz12,Yilmaz14}. 

Existing methods employ level-triggered sampling to report a scalar local process to the FC. Using a similar procedure to report each distinct entry of $\bar{\mU}_t^k$ and $\bar{V}_t^k$ we need $O(n^2)$ parallel procedures, which may be prohibitive in a decentralized setup for large $n$. Hence, we propose to use the approximations introduced in the previous subsection, achieving linear complexity $O(n)$. Moreover, for highly accurate approximations, existing methods transmit multiple bits of information per sample to overcome the overshoot problem, which again can be cumbersome even with $O(n)$ parallel procedures. To that end, we propose an alternative way to handle the overshoot problem. Particularly, in the proposed decentralized estimator, the overshoot in each sample is encoded in time by transmitting a single pulse with very short duration, which greatly helps comply with the stringent energy constraints. 

We will next describe the proposed decentralized estimator based on level-triggered sampling in which each sensor non-uniformly samples the local processes $\{\mD_t^k\}_t$ and $\{\bar{V}_t^k\}_t$, transmits a single pulse for each sample to the FC, and the FC computes $\{\widetilde{\mD}_t\}$ and $\{\widetilde{V}_t\}$ using received information.

\subsubsection{Sampling and Recovery of $\mD_t^k$}

Each sensor $k$ samples each entry $d_{t,i}^k$ of $\mD_t^k$ at a sequence of random times $\{s_{m,i}^k\}_m$ \footnote{The subscript $m$ in the set notation denotes $m\in\bN$.} given by
\be
\label{eq:samp_time_dec}
	s_{m,i}^k \triangleq \min\left\{ t\in\bN: d_{t,i}^k-d_{s_{m-1,i}^k,i}^k \geq \Delta_i^k \right\},~s_{0,i}^k=0,
\ee
where $d_{t,i}^k=\sum_{p=1}^t \frac{(h_{p,i}^k)^2}{\sigma_k^2},~d_{0,i}^k=0$ and $\Delta_i^k>0$ is a constant threshold that controls the average sampling interval. Note that the sampling times $\{s_{m,i}^k\}_m$ in \eqref{eq:samp_time_dec} are dynamically determined by the signal to be sampled, i.e., realizations of $d_{t,i}^k$. Hence, they are random, whereas sampling times in the conventional uniform sampling are deterministic with a certain period. According to the sampling rule in \eqref{eq:samp_time_dec}, a sample is taken whenever the signal level $d_{t,i}^k$ increases by at least $\Delta_i^k$ since the last sampling time. Note that $d_{t,i}^k=\sum_{p=1}^t \frac{(h_{p,i}^k)^2}{\sigma_k^2}$ is non-decreasing in $t$.

\begin{figure}
\centering
\includegraphics[scale=0.45]{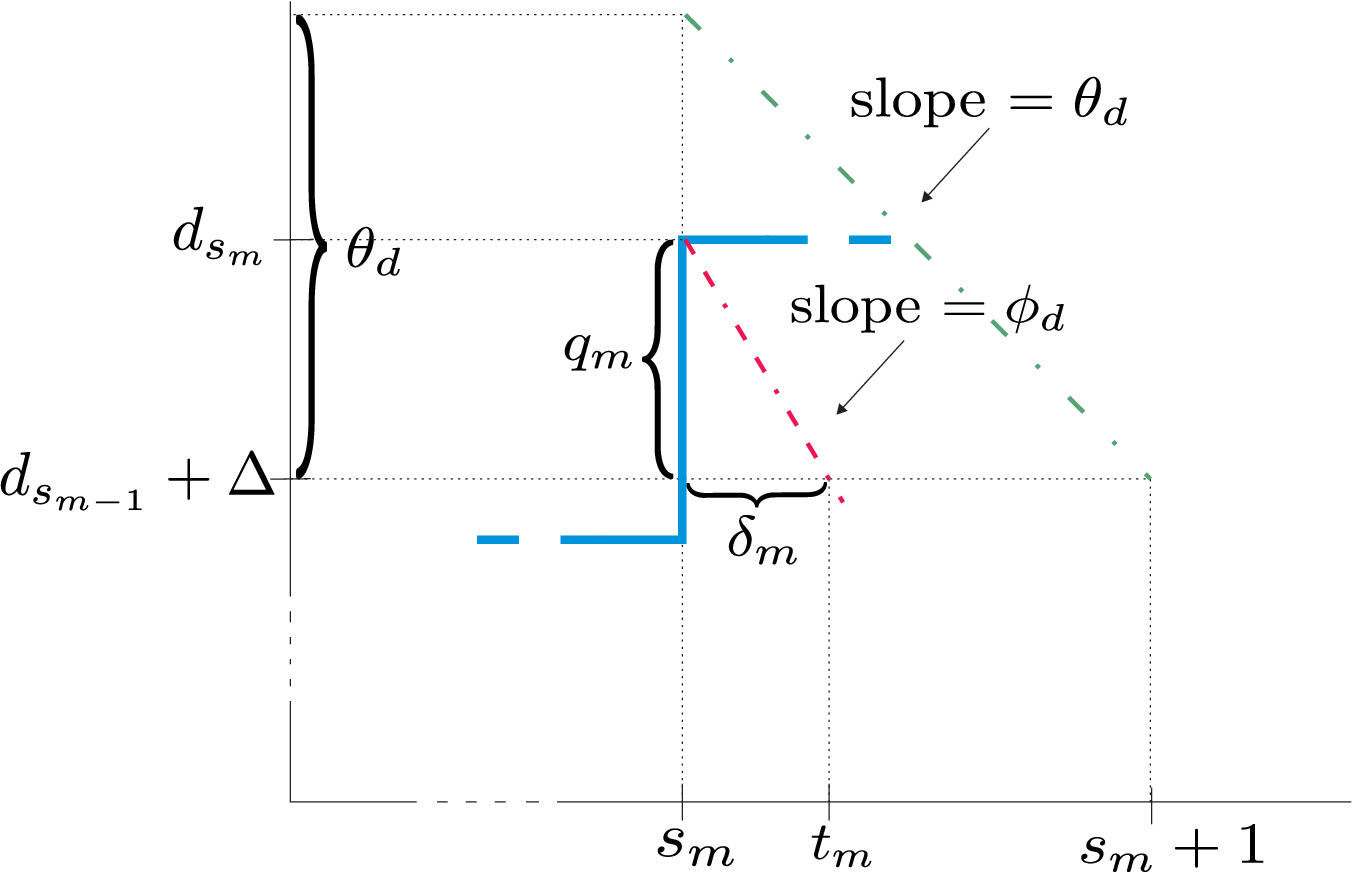}
\caption{Illustration of sampling time $s_m$, transmission time $t_m$, transmission delay $\delta_m$ and overshoot $q_m$. We encode $q_m=(d_{s_m}-d_{s_{m-1}})-\Delta<\theta_d$ in $\delta_m=t_m-s_m<1$ using the slope $\phi_d>\theta_d$.}
\label{fig:over}
\end{figure}

At each sampling time $s_{m,i}^k$, sensor $k$ transmits a single pulse to the FC at time $t_{m,i}^k \triangleq s_{m,i}^k+\delta_{m,i}^k$, indicating that $d_{t,i}^k$ has increased by at least $\Delta_i^k$ since the last sampling time $s_{m-1,i}^k$. The delay $\delta_{m,i}^k$ between the transmission time and the sampling time is used to linearly encode the overshoot
\be
\label{eq:over}
    q_{m,i}^k \triangleq \left(d_{s_{m,i}^k,i}^k-d_{s_{m-1,i}^k,i}^k\right)-\Delta_i^k,
\ee
and given by
\be
\label{eq:ov_enc}
    \delta_{m,i}^k = \frac{q_{m,i}^k}{\phi_d}\in[0,1),
\ee
where $\phi_d^{-1}$ is the slope of the linear encoding function, as shown in Fig. \ref{fig:over}, known to sensors and the FC.

Assume a global clock, that is, the time index $t\in\bN$ is the same for all sensors and the FC, meaning that the FC knows the potential sampling times. Assume further ultra-wideband (UWB) channels between sensors and the FC, in which the FC can determine the time of flight of pulses transmitted from sensors. Then, FC can measure the transmission delay $\delta_{m,i}^k$ if it is bounded by unit time, i.e.,  $\delta_{m,i}^k \in [0,1)$. To ensure this, from \eqref{eq:ov_enc}, we need to have $\phi_d>q_{m,i}^k,~\forall k,m,i$. Assuming a bound for overshoots, i.e., $q_{m,i}^k<\theta_d,~\forall k,m,i$, we can achieve this by setting $\phi_d>\theta_d$.

Consequently, the FC can uniquely decode the overshoot by computing $q_{m,i}^k=\phi_d \delta_{m,i}^k$ (cf. Fig. \ref{fig:over}), using which it can also find the increment occurred in $d_{t,i}^k$ during the interval $(s_{m-1,i}^k,s_{m,i}^k]$ as $d_{s_{m,i}^k,i}^k-d_{s_{m-1,i}^k,i}^k=\Delta_i^k+q_{m,i}^k$ from \eqref{eq:over}. It is then possible to reach the signal level $d_{s_{m,i}^k,i}^k$ by accumulating the increments occurred until the $m$-th sampling time, i.e.,
\be
\label{eq:sig_rec}
    d_{s_{m,i}^k,i}^k = \sum_{\ell=1}^m \left( \Delta_i^k+q_{\ell,i}^k \right) = m\Delta_i^k + \sum_{\ell=1}^m q_{\ell,i}^k.
\ee
Using $\big\{ d_{s_{m,i}^k,i}^k \big\}_m$ the FC computes the staircase approximation $\widetilde{d}_{t,i}^k$ as
\be
\label{eq:d_approx}
	\widetilde{d}_{t,i}^k = d_{s_{m,i}^k,i}^k, ~t\in[t_{m,i}^k,t_{m+1,i}^k),
\ee
which is updated when a new pulse is received from sensor $k$, otherwise kept constant. Such approximate local signals of different sensors are next combined to obtain the approximate global signal $\widetilde{d}_{t,i}$ as
\be
\label{eq:d_app_glo}
	\widetilde{d}_{t,i} = \sum_{k=1}^K \widetilde{d}_{t,i}^k.
\ee
In practice, when the $m$-th pulse in the global order regarding dimension $i$ is received from sensor $k_m$ at time $t_{m,i}$, instead of computing \eqref{eq:sig_rec}--\eqref{eq:d_app_glo} the FC only updates $\widetilde{d}_{t,i}$ as
\be
\label{eq:d_app_upd}
	\widetilde{d}_{t_{m,i},i} = \widetilde{d}_{t_{m-1,i},i} + \Delta_i^{k_m} + q_{m,i},~\widetilde{d}_{0,i}=\epsilon,
\ee
and keeps it constant when no pulse arrives. We initialize $\widetilde{d}_{t,i}$ to a small constant $\epsilon$ to prevent dividing by zero while computing the test statistic [cf. \eqref{eq:tr_upd}].

Note that in general $\widetilde{d}_{t_{m,i},i} \not= d_{s_{m,i},i}$ unlike \eqref{eq:d_approx} since all sensors do not necessarily sample and transmit at the same time. The approximations $\big\{\widetilde{d}_{t,i}\big\}_i$ form $\widetilde{\mD}_t=\text{diag}(\widetilde{d}_{t,1},\ldots,\widetilde{d}_{t,n})$, which is used in \eqref{eq:stop_time_dec2} and \eqref{eq:est_dec} to compute the stopping time and the estimator, respectively. Note that to determine the stopping time as in \eqref{eq:stop_time_dec2} we need to compute $\Tr\left(\widetilde{\mU}_t^{-1}\right)$ using \eqref{eq:dec_approx1} at times $\big\{t_m\big\}$ when a pulse is received from any sensor regarding any dimension. Fortunately, when the $m$-th pulse in the global order is received from sensor $k_m$ at time $t_m$ regarding dimension $i_m$ we can compute $\Tr\left(\widetilde{\mU}_{t_m}^{-1}\right)$ recursively as follows
\be
\label{eq:tr_upd}
    \Tr\left(\widetilde{\mU}_{t_m}^{-1}\right) = \Tr\left(\widetilde{\mU}_{t_{m-1}}^{-1}\right) - \frac{\kappa_{i_m}(\Delta_{i_m}^{k_m} + q_m)}{\widetilde{d}_{t_m,i_m} \widetilde{d}_{t_{m-1},i_m}},~~\Tr\left(\widetilde{\mU}_0^{-1}\right) = \sum_{i=1}^n \frac{\kappa_i}{\epsilon},
\ee
where $\kappa_i$ is the $i$-th diagonal element of the inverse correlation matrix $\mR^{-1}$, known to the FC. In \eqref{eq:tr_upd} pulse arrival times are assumed to be distinct for the sake of simplicity. In case multiple pulses arrive at the same time, the update rule will be similar to \eqref{eq:tr_upd} except that it will consider all new arrivals together.

\subsubsection{Sampling and Recovery of $\bar{V}_t^k$}

Similar to \eqref{eq:samp_time_dec} each sensor $k$ samples each entry $\bar{v}_{t,i}^k$ of $\bar{V}_t^k$ at a sequence of random times $\big\{\alpha_{m,i}^k\big\}_m$ written as
\be
\label{eq:samp_time_dec_v}
	\alpha_{m,i}^k \triangleq \min\left\{ t\in\bN: \big|\bar{v}_{t,i}^k-\bar{v}_{\alpha_{m-1,i}^k,i}^k\big| \geq \gamma_i^k \right\},~\alpha_{0,i}^k=0,
\ee
where $\bar{v}_{t,i}^k=\sum_{p=1}^t \frac{h_{p,i}^k y_{p}^k}{\sigma_k^2}$ and $\gamma_i^k$ is a constant threshold, available to both sensor $k$ and the FC. It has been shown in \cite[Section IV-B]{Yilmaz12} that $\gamma_i^k=\gamma_i$ can be determined by
\be
    \gamma_i \tanh\left(\frac{\gamma_i}{2}\right) = \frac{1}{R} \sum_{k=1}^K |\Exp[\bar{v}_{1,i}^k]|
\ee
to ensure that the FC receives messages with an average rate of $R$ messages per unit time interval. Since $\bar{v}_{t,i}^k$ is neither increasing nor decreasing, we use two thresholds $\gamma_i^k$ and $-\gamma_i^k$ in the sampling rule given in \eqref{eq:samp_time_dec_v}. Specifically, a sample is taken whenever $\bar{v}_{t,i}^k$ increases or decreases by at least $\gamma_i^k$ since the last sampling time. Then, sensor $k$ at time $p_{m,i}^k \triangleq \alpha_{m,i}^k+\beta_{m,i}^k$ transmits a single pulse $b_{m,i}^k$ to the FC, indicating whether $\bar{v}_{t,i}^k$ has changed by at least $\gamma_i^k$ or $-\gamma_i^k$ since the last sampling time $\alpha_{m-1,i}^k$. We can simply write $b_{m,i}^k$ as
\be
\label{eq:bit_dec}
    b_{m,i}^k = \text{sign}\big(\bar{v}_{\alpha_{m,i}^k,i}^k-\bar{v}_{\alpha_{m-1,i}^k,i}^k\big),
\ee
where $b_{m,i}^k=1$ implies that $\bar{v}_{\alpha_{m,i}^k,i}^k-\bar{v}_{\alpha_{m-1,i}^k,i}^k \geq \gamma_i^k$ and $b_{m,i}^k=-1$ indicates that $\bar{v}_{\alpha_{m,i}^k,i}^k-\bar{v}_{\alpha_{m-1,i}^k,i}^k \leq -\gamma_i^k$. The overshoot $\eta_{m,i}^k \triangleq \big|\bar{v}_{\alpha_{m,i}^k,i}^k-\bar{v}_{\alpha_{m-1,i}^k,i}^k\big| - \gamma_i^k$ is linearly encoded in the transmission delay as before. Similar to \eqref{eq:ov_enc} the transmission delay is written as $\beta_{m,i}^k=\frac{\eta_{m,i}^k}{\phi_v}$, where $\phi_v^{-1}$ is the slope of the encoding function, available to sensors and the FC.
\begin{algorithm}[tb]\small
\caption{\small The level-triggered sampling procedure at the $k$-th sensor for the $i$-th dimension}
\label{alg:sen_alg}
\baselineskip=0.5cm
\begin{algorithmic}[1]
\STATE Initialization: $t \gets 0, \ \; m \gets 0, \ \; \ell \gets 0, \ \; \chi \gets 0, \ \; \psi \gets 0$
\WHILE {$\chi < \Delta_i^k$ \AND $\psi \in (-\gamma_i^k,\gamma_i^k)$}
    \STATE $t \gets t+1$
    \STATE $\chi \gets \chi + \frac{(h_{t,i}^k)^2}{\sigma_k^2}$
    \STATE $\psi \gets \psi + \frac{h_{t,i}^k y_t^k}{\sigma_k^2}$
\ENDWHILE
\IF {$\chi \geq \Delta_i^k$ \COMMENT{sample $d_{t,i}^k$}}
    \STATE $m \gets m+1$
    \STATE $s_{m,i}^k=t$
    \STATE Send a pulse to the fusion center at time instant $t_{m,i}^k=s_{m,i}^k+\frac{\chi-\Delta_i^k}{\phi_d}$
    \STATE $\chi \gets 0$
\ENDIF
\IF {$\psi \not\in (-\gamma_i^k,\gamma_i^k)$ \COMMENT{sample $\bar{v}_{t,i}^k$}}
    \STATE $\ell \gets \ell+1$
    \STATE $\alpha_{\ell,i}^k=t$
    \STATE Send $b_{\ell,i}^k = {\rm sign}(\psi)$ to the fusion center at time instant $p_{\ell,i}^k=\alpha_{\ell,i}^k+\frac{|\psi|-\gamma_i^k}{\phi_v}$
    \STATE $\psi \gets 0$
\ENDIF
\STATE Stop if the fusion center instructs so; otherwise go to line 2.
\end{algorithmic}
\end{algorithm}

Assume again that (i) there exists a global clock among sensors and the FC, (ii) the FC determines channel delay (i.e., time of flight), and (iii) overshoots are bounded by a constant, i.e., $\eta_{m,i}^k < \theta_v, ~\forall k,m,i$, and we set $\phi_v>\theta_v$. With these assumptions we ensure that the FC can measure the transmission delay $\beta_{m,i}^k$, and accordingly decode the overshoot as $\eta_{m,i}^k=\phi_v \beta_{m,i}^k$. Then, upon receiving the $m$-th pulse $b_{m,i}$ regarding dimension $i$ from sensor $k_m$ at time $p_{m,i}$ the FC performs the following update,
\be
\label{eq:d_app_upd_v}
	\widetilde{v}_{p_{m,i},i} = \widetilde{v}_{p_{m-1,i},i} + b_{m,i} \big(\gamma_i^{k_m} + \eta_{m,i}\big),
\ee
where $\big\{\widetilde{v}_{t,i}\big\}_i$ compose the approximation $\widetilde{V}_t=[\widetilde{v}_{t,1},\ldots,\widetilde{v}_{t,n}]^T$. Recall that the FC employs $\widetilde{V}_t$ to compute the estimator as in \eqref{eq:est_dec}.

\begin{algorithm}[tb]\small
\caption{\small The sequential estimation procedure at the fusion center}
\label{alg:FC_alg}
\baselineskip=0.5cm
\begin{algorithmic}[1]
\STATE Initialization:   $\Tr\gets \sum_{i=1}^n\frac{\kappa_i}{\epsilon}, \ \; m \gets 1, \ \; \ell\gets1, \ \; \widetilde{d}_i \gets \epsilon~\forall i, \ \; \widetilde{v}_i \gets 0~\forall i$
\WHILE {$\Tr < \widetilde{C}$}
    \STATE  Wait to receive a pulse
    \IF {$m$-th pulse about $d_{t,i}$ arrives from sensor $k$ at time $t$}
        \STATE $q_m=\phi_d(t-\lfloor t \rfloor)$
        \STATE $\Tr \gets \Tr - \frac{\kappa_i(\Delta_i^k+q_m)}{\widetilde{d}_i(\widetilde{d}_i+\Delta_i^k+q_m)}$
        \STATE $\widetilde{d}_i=\widetilde{d}_i+\Delta_i^k+q_m$
        \STATE $m \gets m+1$
    \ENDIF
    \IF {$\ell$-th pulse $b_{\ell}$ about $v_{t,j}$ arrives from sensor $k$ at time $t$}
        \STATE $\eta_{\ell}=\phi_v(t-\lfloor t \rfloor)$
        \STATE $\widetilde{v}_j=\widetilde{v}_j+b_{\ell}(\gamma_j^k+\eta_{\ell})$
        \STATE $\ell \gets \ell+1$
    \ENDIF
\ENDWHILE
\STATE Stop at time $\widetilde{\cT}=t$
\STATE $\widetilde{\mD}=\text{diag}(\widetilde{d}_1,\ldots,\widetilde{d}_n), \ \; \widetilde{\mU}^{-1}=\widetilde{\mD}^{-1/2}\mR^{-1}\widetilde{\mD}^{-1/2}, \ \; \widetilde{V}=[\widetilde{v}_1,\ldots,\widetilde{v}_n]^T$
\STATE $\widetilde{X}=\widetilde{\mU}^{-1}\widetilde{V}$
\STATE Instruct sensors to stop
\end{algorithmic}
\end{algorithm}

The level-triggered sampling procedure at each sensor $k$ for each dimension $i$ is summarized in Algorithm \ref{alg:sen_alg}. Each sensor $k$ runs $n$ of these procedures in parallel. The sequential estimation procedure at the FC is also summarized in Algorithm \ref{alg:FC_alg}. We assumed, for the sake of clarity, that each sensor transmits pulses to the FC for each dimension through a separate channel, i.e., parallel architecture. On the other hand, in practice the number of parallel channels can be decreased to two by using identical sampling thresholds  $\Delta$ and $\gamma$ for all sensors and for all dimensions in \eqref{eq:samp_time_dec} and \eqref{eq:samp_time_dec_v}, respectively. Moreover, sensors can even employ a single channel to convey information about local processes $\{d_{t,i}^k\}$ and $\{\bar{v}_{t,i}^k\}$ by sending ternary digits to the FC. This is possible since pulses transmitted for $\{d_{t,i}^k\}$ are unsigned.

\subsection{Discussions}
\label{sec:disc}

We introduced the decentralized estimator in Section \ref{sec:dest} initially for a continuous-time system with infinite precision. In practice, due to bandwidth constraints, discrete-time systems with finite precision are of interest. For example, in such systems, the overshoot $q_{m,i}^k\in\left[j\frac{\theta_d}{N},(j+1)\frac{\theta_d}{N}\right),j=0,1,\ldots,N-1,$ is quantized into $\hat{q}_{m,i}^k=\left(j+\frac{1}{2}\right)\frac{\theta_d}{N}$ where $N$ is the number of quantization levels. More specifically, a pulse is transmitted at time $t_{m,i}^k=s_{m,i}^k+\frac{j+1/2}{N}$, where the transmission delay $\frac{j+1/2}{N}\in(0,1)$ encodes $\hat{q}_{m,i}^k$. This transmission scheme is called pulse position modulation (PPM).

In UWB and optical communication systems, PPM is effectively employed. In such systems, $N$, which denotes the precision, can be easily made large enough so that the quantization error $|\hat{q}_{m,i}^k-q_{m,i}^k|$ becomes insignificant. Compared to conventional transmission techniques which convey information by varying the power level, frequency, and/or phase of a sinusoidal wave, PPM (with UWB) is extremely energy efficient at the expense of high bandwidth usage since only a single pulse with very short duration is transmitted per sample. Hence, PPM suits well to energy-constrained sensor network systems.

\subsection{Simulation Results}
\label{sec:sim}

\begin{figure}[htb]
\centering
\includegraphics[scale=0.7]{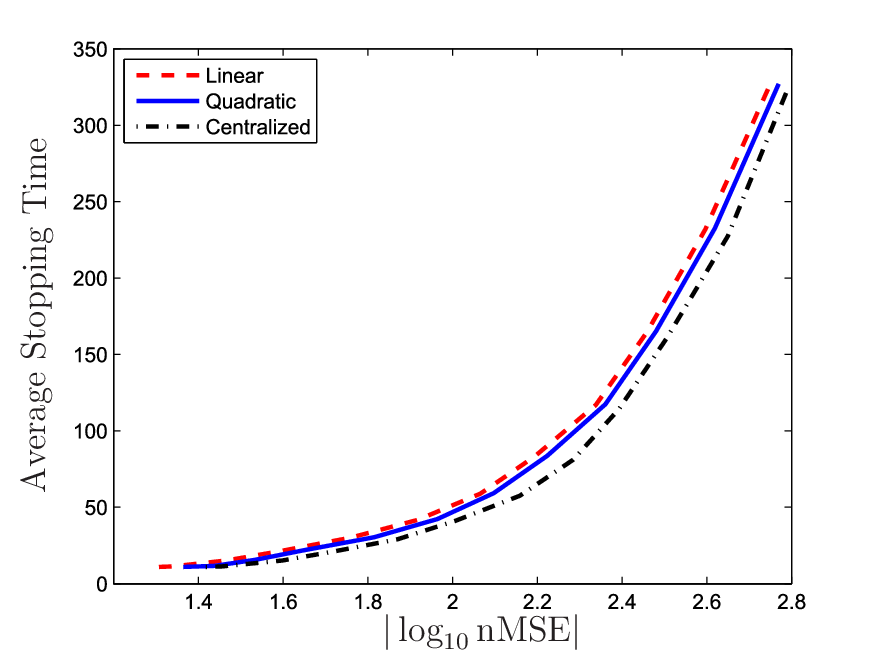}
\caption{Average stopping time performances of the optimal centralized scheme and the decentralized schemes based on level-triggered sampling with quadratic and linear complexity vs. normalized MSE values when scaling coefficients are uncorrelated, i.e., $r_{ij}=0,\forall i,j$.}
\label{fig:mse_unc}
\end{figure}

\begin{figure}[htb]
\centering
\includegraphics[scale=0.7]{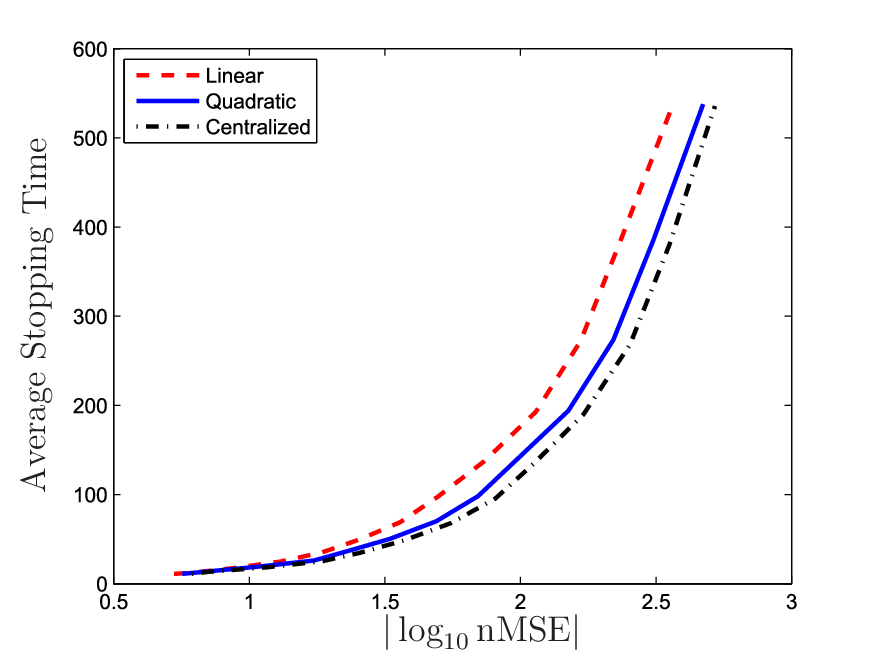}
\caption{Average stopping time performances of the optimal centralized scheme and the decentralized schemes based on level-triggered sampling with quadratic and linear complexity vs. normalized MSE values when scaling coefficients are correlated with $r_{ij}=0.5,\forall i,j$.}
\label{fig:mse_cor}
\end{figure}

We next provide simulation results to compare the performances of the proposed scheme with linear complexity, given in Algorithm \ref{alg:sen_alg} and Algorithm \ref{alg:FC_alg}, the unsimplified version of the proposed scheme with quadratic complexity and the optimal centralized scheme. A wireless sensor network with $10$ identical sensors and an FC is considered to estimate a $5$-dimensional deterministic vector of parameters, i.e., $n=5$. We assume i.i.d. Gaussian noise with unit variance at all sensors, i.e., $w_t^k\sim\cN(0,1),\forall k,t$. We set the correlation coefficients $\{r_{ij}\}$ [cf. \eqref{eq:uncor3}] of the vector $H_t^k$ to $0$ in Fig. \ref{fig:mse_unc} and $0.5$ in Fig. \ref{fig:mse_cor} to test the performance of the proposed scheme in the uncorrelated and correlated cases, respectively. We compare the average stopping time performance of the proposed scheme with linear complexity to those of the other two schemes for different MSE values. In Fig. \ref{fig:mse_unc} and Fig. \ref{fig:mse_cor}, the horizontal axis represents the signal-to-error ratio in dB, where $\text{nMSE} \triangleq \frac{\text{MSE}}{\|X\|_2^2}$, i.e., the MSE normalized by the square of the Euclidean norm of the vector to be estimated.

In the uncorrelated case, where $r_{ij}=0,~\forall i,j,~i\not=j$, the proposed scheme with linear complexity nearly attains the performance of the unsimplified scheme with quadratic complexity as seen in Fig. \ref{fig:mse_unc}. This result is rather expected since in this case $\bar{\mU}_t \cong \mD_t$ for sufficiently large $t$, where $\bar{\mU}_t$ and $\mD_t$ are  used to compute the stopping time and the estimator in the unsimplified and simplified schemes, respectively. Strikingly the decentralized schemes (simplified and unsimplified) achieve very close performances to that of the optimal centralized scheme, which is obviously unattainable in a decentralized system, thanks to the efficient information transmission through level-triggered sampling.
It is seen in Fig. \ref{fig:mse_cor} that the proposed simplified scheme exhibits an average stopping time performance close to those of the unsimplified scheme and the optimal centralized scheme even when the scaling coefficients $\{h_{t,i}^k\}_i$ are correlated with $r_{ij}=0.5,~\forall i,j,~i\not=j$, justifying the simplification proposed in Section \ref{sec:simp} to obtain linear complexity.
\begin{figure}[htb]
\centering
\includegraphics[scale=0.7]{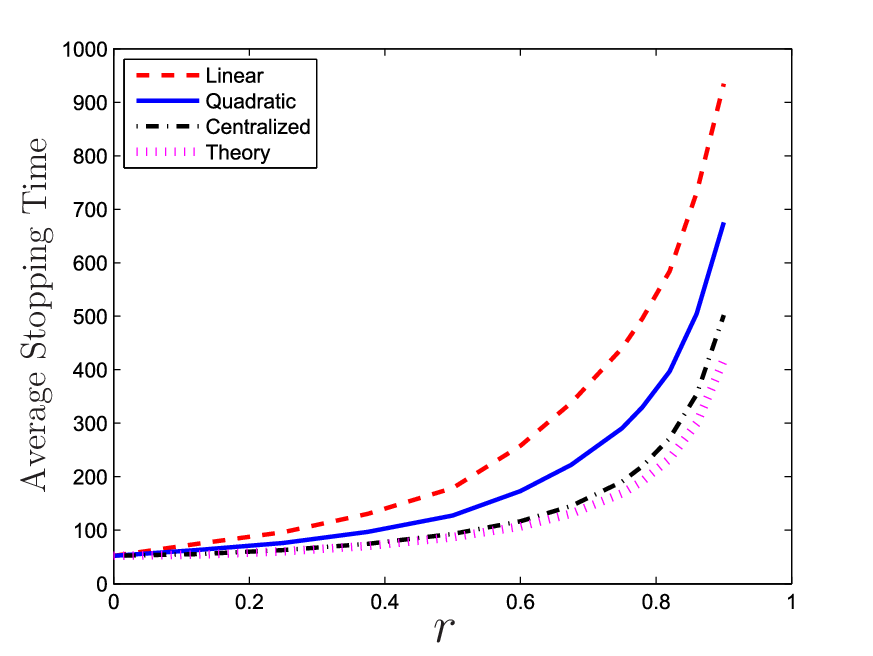}
\caption{Average stopping time performances of the optimal centralized scheme and the decentralized schemes based on level-triggered sampling with quadratic and linear complexity vs. correlation coefficient for normalized MSE fixed to $10^{-2}$.}
\label{fig:rho}
\end{figure}

Finally, in Fig. \ref{fig:rho} we fix the normalized MSE value at $10^{-2}$ and plot average stopping time against the correlation coefficient $r$ where $r_{ij}=r,~\forall i,j,~i\not=j$.
We observe an exponential growth in average stopping time of each scheme as $r$ increases. The average stopping time of each scheme becomes infinite at $r=1$ since in this case only some multiples of a certain linear combination of the parameters to be estimated, i.e., $h_{t,1}^k \sum_{i=1}^n c_i x_i$, are observed under the noise $w_t^k$ at each sensor $k$ at each time $t$, hence it is not possible to recover the individual parameters. Specifically, it can be shown that $c_i=\sqrt{\frac{\Exp\big[\big(h_{t,i}^k\big)^2\big]}{\Exp\big[\big(h_{t,1}^k\big)^2\big]}}$, which is the same for all sensors as we assume identical sensors. To see the mechanism that causes the exponential growth consider the computation of $\Tr(\bar{\mU}_t^{-1})$, which is used to determine the stopping time in the optimal centralized scheme. From \eqref{eq:dec_approx1} we write
\be
\label{eq:exp_grw}
  \Tr(\bar{\mU}_t^{-1}) \cong \Tr(\mD_t^{-1}\mR^{-1}) = \sum_{i=1}^n \frac{\kappa_i}{d_{t,i}}
\ee
for sufficiently large $t$, where $d_{t,i}$ and $\kappa_i$ are the $i$-th diagonal elements of the matrices $\mD_t$ and $\mR^{-1}$, respectively. For instance, we have $\kappa_i=1,\forall i$, $\kappa_i=8.0435,\forall i$ and $\kappa_i=\infty$ when $r=0$, $r=0.9$ and $r=1$, respectively. Assuming that the scaling coefficients have the same mean and variance when $r=0$ and $r=0.9$, we have similar $d_{t,i}$ values [cf. \eqref{eq:diag}] in \eqref{eq:exp_grw}, hence the stopping time of $r=0.9$ is approximately $8$ times that of $r=0$ for the same accuracy level. Since MSE $=\Exp\big[\|\hat{\X}_{\cT}-X\|_2^2\big]=\Tr(\bar{\mU}_t^{-1})$ in the centralized scheme, using $\kappa_i$ for different $r$ values we can approximately know how the average stopping time changes as $r$ increases for a given MSE value. As shown in Fig. \ref{fig:rho} with the label ``Theory" this theoretical curve is in a good match with the numerical result. The small discrepancy at high $r$ values is due to the high sensitivity of the WLS estimator in \eqref{eq:WLS1_dec} to numerical errors when the stopping time is large. The high sensitivity is due to multiplying the matrix $\bar{\mU}_{\cT}^{-1}$ with very small entries by the vector $\bar{V}_{\cT}$ with very large entries while computing the estimator $\hat{\X}_{\cT}$ in \eqref{eq:WLS1_dec} for a large $\cT$. The decentralized schemes suffer from a similar high sensitivity problem [cf. \eqref{eq:est_dec}] much more than the centralized scheme since making error is inherent in a decentralized system. Moreover, in the decentralized schemes the MSE is not given by the stopping time statistic $\Tr\big(\widetilde{\mU}_t^{-1}\big)$, hence ``Theory" does not match well the curves for the decentralized schemes. Although it cannot be used to estimate the rates of the exponential growths of the decentralized schemes, it is still useful to explain the mechanism behind them as the decentralized schemes are derived from the centralized scheme.

To summarize, with identical sensors any estimator (centralized or decentralized) experiences an exponential growth in its average stopping time as the correlation between scaling coefficients increases since in the extreme case of full correlation, i.e., $r=1$, each sensor $k$, at each time $t$, observes a noisy sample of the linear combination $\sum_{i=1}^n x_i \sqrt{\frac{\Exp\big[\big(h_{t,i}^k\big)^2\big]}{\Exp\big[\big(h_{t,1}^k\big)^2\big]}}$, and thus the stopping time is infinite. As a result of exponentially growing stopping time, the WLS estimator, which is the optimum estimator in our case, i.e., the MVUE, and the decentralized estimators derived from it become highly sensitive to errors as $r$ increases. In either uncorrelated or mildly correlated cases, which are of practical importance, the proposed decentralized scheme with linear complexity performs very close to the optimal centralized scheme as shown in Fig. \ref{fig:mse_unc} and Fig. \ref{fig:mse_cor}, respectively.

\section{Conclusions}
\label{sec:conc}

We have considered the problem of sequential vector parameter estimation under both centralized and decentralized settings. In the centralized setting, we have first sought the optimum sequential estimator under the classical formulation of the problem in which expected stopping time is minimized subject to a constraint on a function of the estimator covariance. Treating the problem with optimal stopping theory we have showed that the optimum solution is intractable for even moderate number of parameters to be estimated. Then, we have considered an alternative formulation that is conditional on the observed regressors, and showed that it has a simple optimum solution for any number of parameters. Using the tractable optimum sequential estimator of the conditional formulation we have also developed a computation- and energy-efficient decentralized estimator. In the decentralized setup, to satisfy the stringent energy constraints we have proposed two novelties in the level-triggered sampling procedure, which is a non-uniform sampling technique. Finally, numerical results have demonstrated that the proposed decentralized estimator has a similar average stopping time performance to that of the optimum centralized estimator.

\section*{Appendix: Proof of Lemma \ref{lem:unc_sca}}

We will first prove that if $\cV(z)$ is non-decreasing, concave and bounded, then so is $G(z)=1+\Exp\left[\cV\left(\frac{z}{1+z h_1^2}\right)\right]$. That is, assume $\cV(z)$ satisfies: (a) $\frac{\text{d}}{\text{d} z}\cV(z)\geq0$, (b) $\frac{\text{d}^2}{\text{d} z^2}\cV(z)<0$, (c) $\cV(z)<c<\infty,\forall z$. Then by (c) we have
  \be
    1+\cV\left(\frac{z}{1+z h_1^2}\right)<1+c,\forall z,
  \ee
  hence $G(z)<1+c$ is bounded. Moreover,
  \be
    \frac{\text{d}}{\text{d} z}\cV\left(\frac{z}{1+z h_1^2}\right) = \frac{\frac{\text{d}}{\text{d} z}\cV(z)}{(1+z h_1^2)^2}>0,~\forall z
  \ee
  by (a), and thus $G(z)$ is non-decreasing. Furthermore,
  \be
    \frac{\text{d}^2}{\text{d} z^2}G(z)=\Exp\Bigg[\frac{\text{d}^2}{\text{d} z^2}\cV\left(\frac{z}{1+z h_1^2}\right)\Bigg] = \Exp\Bigg[ \underbrace{\frac{\frac{\text{d}^2}{\text{d} z^2}\cV(z)}{(1+z h_1^2)^4}}_{<0~\text{by (b)}} + \underbrace{\frac{\frac{\text{d}}{\text{d} z}\cV(z)}{-(1+z h_1^2)^3/2h_1^2}}_{<0~\text{by (a) \& }z=\frac{1}{u}>0}\Bigg],~\forall z,
  \ee
  hence $G(z)$ is concave, concluding the first part of the proof.

\begin{figure}
\centering
\includegraphics[scale=0.35]{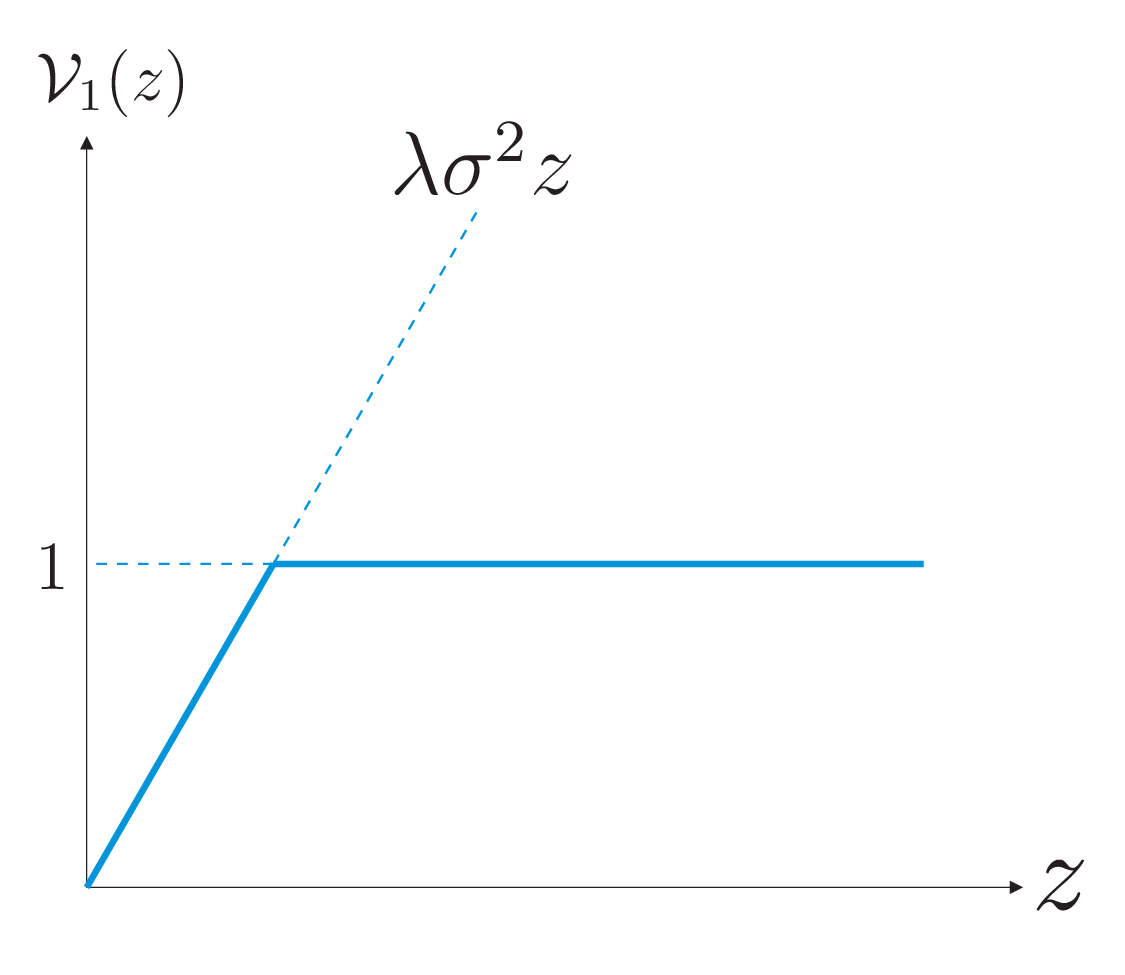}
\caption{The function $\cV_1(z)$ is non-decreasing and concave.}
\label{fig:V_1}
\end{figure}

 Now, it is sufficient to show that $\cV(z)$ is non-decreasing, concave and bounded.
Assume that the limit $\lim_{m\to\infty}\cV_m(z)=\cV(z)$ exists. We will prove the existence of the limit later. First, we will show that $\cV(z)$ is non-decreasing and concave by iterating the functions $\{\cV_m(z)\}$. Start with $\cV_0(z)=0$. Then,
\be
\label{eq:V_1}
    \cV_1(z)=\min\left\{ \lambda \sigma^2 z,1+\Exp\left[\cV_0\left(\frac{z}{1+z h_1^2}\right)\right] \right\}=\min\{ \lambda \sigma^2 z,1\},
\ee
which is non-decreasing and concave as shown in Fig. \ref{fig:V_1}. Similarly we write
\be
    \cV_2(z)=\min\left\{ \lambda \sigma^2 z,1+\Exp\left[\cV_1\left(\frac{z}{1+z h_1^2}\right)\right] \right\},
\ee
where $1+\Exp\left[\cV_1\left(\frac{z}{1+z h_1^2}\right)\right]$ is non-decreasing and concave since $\cV_1(z)$ is non-decreasing and concave.
Hence, $\cV_2(z)$ is non-decreasing and concave since pointwise minimum of non-decreasing and concave functions is again non-decreasing and concave. We can show in the same way that $\cV_m(z)$ is non-decreasing and concave for $m>2$, i.e., $\cV(z)=\cV_{\infty}(z)$ is non-decreasing and concave.

Next, we will show that $\cV(z)$ is bounded. Assume that
\be
\label{eq:assum}
    \cV(z)<\min\{ \lambda\sigma^2 z,c \}=\lambda\sigma^2 z \ind{\lambda\sigma^2 z\leq c} + c\ind{\lambda\sigma^2 z>c}.
\ee
Then, from the definition of $\cV(z)$ we have $1+\Exp\left[ \cV\left(\frac{z}{1+z h_1^2}\right) \right]<c$. Since $\cV(z)$ is non-decreasing, $\Exp\left[ \cV\left(\frac{z}{1+z h_1^2}\right) \right]\leq \Exp\left[ \cV\left(\frac{1}{h_1^2}\right) \right]$. From \eqref{eq:assum} we can write
\be
\label{eq:bdd}
    1+\Exp\left[ \cV\left(\frac{z}{1+z h_1^2}\right)\right] \leq 1+\Exp\left[ \cV\left(\frac{1}{h_1^2}\right) \right] < 1+\Exp\left[ \frac{\lambda\sigma^2}{h_1^2} \ind{\frac{\lambda\sigma^2}{h_1^2}\leq c} \right] + c~\Pro\left(\frac{\lambda\sigma^2}{h_1^2}>c\right),
\ee
Recalling $1+\Exp\left[ \cV\left(\frac{z}{1+z h_1^2}\right)\right]<c$ we want to find a $c$ such that
\be
\label{eq:c}
    1+\Exp\left[ \frac{\lambda\sigma^2}{h_1^2} \ind{\frac{\lambda\sigma^2}{h_1^2}\leq c} \right] + c~\Pro\left(\frac{\lambda\sigma^2}{h_1^2}>c\right) < c.
\ee
For such a $c$ we have
\begin{align}
    1 &< c~\Pro\left(\frac{\lambda\sigma^2}{h_1^2}\leq c\right) - \Exp\left[ \frac{\lambda\sigma^2}{h_1^2} \ind{\frac{\lambda\sigma^2}{h_1^2}\leq c} \right] \nn\\
    &= \Exp\left[ \left( c-\frac{\lambda\sigma^2}{h_1^2} \right) \ind{\frac{\lambda\sigma^2}{h_1^2}\leq c} \right] = \Exp\left[ \left( c-\frac{\lambda\sigma^2}{h_1^2} \right)^{+} \right],
\end{align}
where $(\cdot)^+$ is the positive part operator. We need to show that there exists a $c$ satisfying $\Exp\left[ \left( c-\frac{\lambda\sigma^2}{h_1^2} \right)^{+} \right]>1$. Note that we can write
\begin{align}
\label{eq:bdd2}
	\Exp\left[ \left( c-\frac{\lambda\sigma^2}{h_1^2} \right)^{+} \right] &\geq \Exp\left[ \left( c-\frac{\lambda\sigma^2}{h_1^2} \right)^{+} \ind{h_1^2>\epsilon} \right] \nn\\
	&> \Exp\left[ \left( c-\frac{\lambda\sigma^2}{\epsilon} \right)^{+} \ind{h_1^2>\epsilon} \right] \nn\\
	&= \left( c-\frac{\lambda\sigma^2}{\epsilon} \right)^{+} \Pro(h_1^2>\epsilon),
\end{align}
where $\left( c-\frac{\lambda\sigma^2}{\epsilon} \right)^{+}\to\infty$ as $c\to\infty$ since $\lambda$ and $\epsilon$ are constants. If $\Pro(h_1^2>\epsilon)>0$, which is always true except the trivial case where $h_1=0$ deterministically, then the desired $c$ exists.

Now, what remains is to justify our initial assumption $\cV(z)<\min\{ \lambda\sigma^2 z,c \}$. We will use induction to show that the assumption holds with the $c$ found above. From \eqref{eq:V_1}, we have $\cV_1(z)=\min\{ \lambda\sigma^2 z,1\}<\min\{ \lambda\sigma^2 z,c \}$ since $c>1$. Then, assume that
\be
\label{eq:V_n-1}
	\cV_{m-1}(z)<\min\{ \lambda\sigma^2 z,c \}=\lambda\sigma^2 z \ind{\lambda\sigma^2 z\leq c} + c\ind{\lambda\sigma^2 z>c}.
\ee
We need to show that $\cV_m(z)<\min\{ \lambda\sigma^2 z,c \}$, where
	$\cV_m(z)=\min\left\{ \lambda \sigma^2 z,1+\Exp\left[\cV_{m-1}\left(\frac{z}{1+z h_1^2}\right)\right] \right\}$.
Note that $1+\Exp\left[\cV_{m-1}\left(\frac{z}{1+z h_1^2}\right)\right] \leq 1+\Exp\left[\cV_{m-1}\left(\frac{1}{h_1^2}\right)\right]$ since $\cV_{m-1}(z)$ is non-decreasing. Similar to \eqref{eq:bdd}, from \eqref{eq:V_n-1} we have
\be
	1+\Exp\left[\cV_{m-1}\left(\frac{1}{h_1^2}\right)\right] < 1+\Exp\left[ \frac{\lambda\sigma^2}{h_1^2} \ind{\frac{\lambda\sigma^2}{h_1^2}\leq c} \right] + c~\Pro\left(\frac{\lambda\sigma^2}{h_1^2}>c\right) < c,
\ee
where the last inequality follows from \eqref{eq:c}. Hence,
\be
\label{eq:bdd3}
	\cV_m(z)<\min\{ \lambda\sigma^2 z,c \}, ~\forall m,
\ee
showing that $\cV(z)<\min\{ \lambda\sigma^2 z,c \}$, which is the assumption in \eqref{eq:assum}.

We showed that $\cV(z)$ is non-decreasing, concave and bounded if it exists, i.e., the limit $\lim_{m\to\infty}\cV_m(z)$ exists. Note that we showed in \eqref{eq:bdd3} that the sequence $\{\cV_m\}$ is bounded. If we also show that $\{\cV_m\}$ is monotonic, e.g., non-decreasing, then $\{\cV_m\}$ converges to a finite limit $\cV(z)$. We will again use induction to show the monotonicity for $\{\cV_m\}$. From \eqref{eq:V_1} we write $\cV_1(z)=\min\{\lambda\sigma^2 z,1\}\geq \cV_0(z)=0$. Assuming $\cV_{m-1}(z)\geq \cV_{m-2}(z)$ we need to show that $\cV_m(z)\geq \cV_{m-1}(z)$. Using their definitions we write $\cV_m(z)=\min\left\{ \lambda \sigma^2 z,1+\Exp\left[\cV_{m-1}\left(\frac{z}{1+z h_1^2}\right)\right] \right\}$ and $\cV_{m-1}(z)=\min\left\{ \lambda \sigma^2 z,1+\Exp\left[\cV_{m-2}\left(\frac{z}{1+z h_1^2}\right)\right] \right\}$. We have $1+\Exp\left[\cV_{m-1}\left(\frac{z}{1+z h_1^2}\right)\right] \geq 1+\Exp\left[\cV_{m-2}\left(\frac{z}{1+z h_1^2}\right)\right]$ due to the assumption $\cV_{m-1}(z)\geq \cV_{m-2}(z)$, hence $\cV_m(z)\geq \cV_{m-1}(z)$.

To conclude, we proved that $\cV_m(z)$ is non-decreasing and bounded in $m$, thus the limit $\cV(z)$ exists, which was also shown to be non-decreasing, concave and bounded. Hence, $G(z)$ is non-decreasing, concave and bounded.


\end{document}